\documentclass[man, floatsintext]{apa7}
\pdfoutput=1
\PassOptionsToPackage{hidelinks}{hyperref}  

\usepackage{mathptmx}
\usepackage{amsmath}
\usepackage{lipsum}
\usepackage{algorithmic}
\usepackage[ruled,vlined]{algorithm2e}
\usepackage[american]{babel}
\usepackage[inkscapeformat=png]{svg}
\usepackage{multirow}
\usepackage{diagbox}
\usepackage{csquotes}
\usepackage{color,soul}
\usepackage{caption}
\usepackage{float}

\usepackage[style=apa, backend=biber]{biblatex}
\DeclareLanguageMapping{american}{american-apa}
\DeclareCaptionLabelSeparator*{spaced}{\\[2ex]}
\title{What Contributes to Affective Polarization in Networked Online Environments? Evidence from an Agent-Based Model}
\shorttitle{Polarization Beyond Echo-Chambers}

\authorsnames[1,{1,2},3]{Narayani Vedam, Subhayan Mukerjee, Prasanta Bhattacharya} \authorsaffiliations{{Department of Communications and New Media, National University of Singapore.}, {Centre for Trusted Internet and Community, National University of Singapore.}, {Institute of High Performance Computing (IHPC), 
    Agency for Science, Technology and Research (A*STAR), 1 Fusionopolis Way, \#16-16 Connexis, Singapore 138632, Republic of Singapore.}}
\authornote{
Dr. Subhayan Mukerjee is the corresponding author. Email: mukerjee@nus.edu.sg.}

\leftheader{Weiss}

\abstract{
Affective polarization, or, inter-party hostility, is increasingly recognized as a pervasive issue in democracies worldwide, posing a threat to social cohesion. The digital media ecosystem, now widely accessible and ever-present, has often been implicated in accelerating this phenomenon. However, the precise causal mechanisms responsible for driving affective polarization have been a subject of extensive debate. While the concept of echo chambers, characterized by individuals ensconced within like-minded groups, bereft of counter-attitudinal content, has long been the prevailing hypothesis, accumulating empirical evidence suggests a more nuanced picture. This study aims to contribute to the ongoing debate by employing an agent-based model to illustrate how affective polarization is either fostered or hindered by individual news consumption and dissemination patterns based on ideological alignment. To achieve this, we parameterize three key aspects: (1) The affective asymmetry of individuals' engagement with in-party versus out-party content, (2) The proportion of in-party members within one's social neighborhood, and (3) The degree of partisan bias among the elites within the population.
Subsequently, we observe macro-level changes in affective polarization within the population under various conditions stipulated by these parameters. This approach allows us to explore the intricate dynamics of affective polarization within digital environments, shedding light on the interplay between individual behaviors, social networks, and information exposure.
}


\begin{document}
\maketitle
\section{Introduction}
Democracy, as a system of governance, is predicated on principles such as representation, participation, and the free exchange of ideas \parencite{dahl2008polyarchy,diamond1999developing,delli2001let}. Central to this exchange is opinion diversity \parencite{sunstein2006infotopia,page1993rational}, which presupposes a degree of convergence among the public \textendash\ a notion embodied in the ideal of a public sphere. However, the nature of this convergence has been a subject of debate. Political scientists have argued that American politics in the 20th century witnessed a decline in the influence of political parties and a perceived lack of polarization, both of which were regarded as potential threats to democratic stability \parencite{fiorina1980decline,APSA1950}. In contrast, the 21st century has witnessed a resurgence of polarization, particularly evident in ideological and affective dimensions \parencite{abramowitz2008polarization,fiorina2011culture,lelkes2016mass,iyengar2019origins}. While research on the extent and nature of issue-based or ideological divisions within the general public has yielded mixed conclusions  \parencite{abramowitz2010disappearing,fiorina2008polarization,enders2021issues}, scholars on both sides of the debate speculate that partisan and ideological identities may be coalescing \parencite{lelkes2016mass,levendusky2009partisan,iyengar2012affect}.

Partisanship has historically played a crucial role in shaping political attitudes and behaviors \parencite{lazarsfeld1954friendship,campbell1980american,johnston2006party,gimpel2003partisan}. However, recent years have witnessed its intensification and subsequent effects on downstream outcomes, notably the increase in affective polarization -- characterized by hostility and antagonism towards perceived political adversaries \parencite{iyengar2012affect}. 
This phenomenon is laden with socio-political implications, and has spurred a wave of scholarly inquiry within the United States \parencite{iyengar2019origins,lelkes2017limits,lelkes2016mass,druckman2022mis,druckman2021affective,kingzette2021you,robison2019group,castle2021partisanship,finkel2020political} and beyond \parencite{gidron2020american,reiljan2020fear,boxell2022cross,garzia2021negative,garzia2023affective,bettarelli2023regional,kekkonen2021affective,hobolt2021divided,neyazi2023political,jost2022cognitive,knudsen2021affective}. Increasing evidence highlights the detrimental effects of affective polarization on democratic stability \parencite{kingzette2021affective}, with spillover effects that permeate various social domains, including familial and marital relationships, book clubs, and even recreational activities such as sports \parencite{dellaposta2015liberals,huber2017political,chen2018effect,filsinger2024asymmetric}. Furthermore, its interpersonal ramifications include heightened reluctance to engage with political opponents \parencite{frimer2017liberals} and a progressively vitriolic tone in political discourse \parencite{mason2018uncivil}.

Numerous studies have previously implicated the digital media ecosystem \textendash\ increasingly ubiquitous and accessible \textendash\ as a catalyst exacerbating this phenomenon \parencite{lelkes2017hostile,allcott2020welfare}. However, the transpiring causal mechanisms -- presence of echo chambers or filter bubbles --  underlying this phenomenon have been a subject of extensive debate. While the creation of echo chambers and filter bubbles devoid of counter-attitudinal content has long served as the primary hypothesis \parencite{sunstein2018republic,garrett2009echo,madsen2018large,ross2022echo}, there is accumulating empirical evidence to suggest otherwise \parencite{figa2022through,dubois2018echo,de2021no,vaccari2016echo,barbera2015tweeting}. Moreover, as \textcite{guess2018avoiding} and \textcite{ross2022echo} argue, echo-chambers and filter bubbles may be overstated in the first place.
Furthermore, elites also exert disproportionate influence in shaping divisive narratives and reinforcing partisan divides, thereby amplifying polarization \parencite{enders2021issues,banda2018elite,druckman2019we}. Despite these insights, much remains unknown about how these dynamics interact to influence the polarization process, especially the affective dimension. Consequently, bridging this gap is essential for developing a nuanced and comprehensive understanding of the factors that exacerbate affective divides.

Our current study provides an empirical framework for addressing key research questions on the mechanisms driving affective polarization in digital media environments. Specifically, we examine how affective asymmetry, partisan neighborhood composition, and elite group dynamics influence the evolution of affective polarization. Our model explores whether cross-cutting exposure mitigates or exacerbates partisan hostility and investigates the role of majoritarianism -- among both the general public and elites -- in amplifying affective divides. Our findings indicate that affective asymmetry plays a central role in accelerating polarization, while ideological balance within the general public paradoxically intensifies partisan divides by facilitating broader news dissemination. Additionally, we find that elite composition significantly shapes ideological reinforcement and deters a rapid rise in affective polarization, with homogeneous elites restricting cross-partisan exposure. These insights contribute to ongoing debates on whether polarization inherently threatens democratic discourse or, under certain conditions, fosters engagement with diverse perspectives.

\subsection{Social Media and Polarization}

The global proliferation of smartphones over the past decade has significantly increased social media usage for news consumption \parencite{newman2024reuters}. Platforms such as Facebook and X (formerly, Twitter) have emerged as dominant sources of news, as evidenced by a recent Reuters survey encompassing thirty-eight countries \parencite{newman2024reuters}. The unique technological features of these platforms \textendash\ such as real-time sharing, algorithmic personalization, and widespread accessibility \textendash\ have fundamentally altered how individuals consume, share, and engage with news \parencite{gonzalez2023social}. This shift has broadened the scope of political discussions, transitioning them from private settings to the public domain and exposing them to more diverse audiences \parencite{tucker2018social,van2017political,lelkes2017hostile}. The digital traces left by online news consumption and dissemination have afforded communication scholars unprecedented opportunities to explore causal mechanisms at both micro and macro levels, alongside their aggregate effects on socio-political outcomes. Researchers have focused particularly on polarization and its various manifestations, highlighting the critical importance of this issue in contemporary political discourse \parencite{lelkes2017hostile,bail2018exposure,cinelli2021echo,boxell2022cross,levy2021social}.

In contrast to the media landscape of the pre-digital era, characterized by a limited number of ostensibly non-partisan news channels, the current media environment is marked by the proliferation of news outlets, both online and offline, each exhibiting varying degrees of partisan bias \parencite{stroud2011niche}. The reach of these outlets has extended beyond discretionary selective exposure, largely owing to social media platforms, resulting in inadvertent consumption of partisan news content even among politically disengaged individuals \parencite{kobayashi2024partisan,fletcher2018people,weeks2017incidental}. Therefore, political communication scholars have sought to investigate partisan news exposure as a key factor in assessing how social media influences attitudes and contributes to either exacerbating or mitigating affective polarization \parencite{kubin2021role}. \textcite{settle2018frenemies} finds that social media platforms, particularly Facebook, can intensify affective polarization, even among users who show little interest in political matters. This is corroborated in large-scale field experiments that have shown that deactivating Facebook for users who frequently consume political news can significantly reduce their levels of political polarization \parencite{allcott2020welfare}. Elsewhere, work by \textcite{overgaard2024perceiving} illustrates that exposure to unifying news content -- that is non-partisan -- on social media can mitigate affective polarization by altering individuals’ meta-perceptions. 

Although some studies \parencite{kubin2021role, settle2018frenemies, allcott2020welfare, overgaard2024perceiving} suggest a positive correlation between social media platforms and affective polarization, research specifically on the causal mechanisms driving affective polarization through social media remains limited. This is in contrast with the extensive literature that examines how social media platforms contribute to ideological or issue polarization. A prominent hypothesis in this literature posits that the information-rich media landscape has facilitated the emergence of ideology-congruent silos, which \textcite{sunstein2018republic} refers to as ``information cocoons.'' These cocoons arise from selective exposure \textendash\ often driven by confirmation bias \textendash\ and algorithmic curation of partisan content. When segregation in the opinion space is reflected in homophilic interactions among users, coupled with biases in information diffusion towards like-minded peers, echo chambers emerge \parencite{garrett2009echo, cinelli2021echo, madsen2018large,interian2023network}. Such ideologically congruent online spaces may contribute to the escalation of ideological and affective polarization due to their pronounced ideological orientation \parencite{ross2022echo}. For instance, \textcite{quattrociocchi2016echo} found that Italian and U.S. Facebook users exhibit behaviors consistent with echo chamber dynamics, leading to increased issue polarization within those groups. In \textcite{wojcieszak2021echo}, the analysis of politically engaged Twitter users revealed a greater tendency to share content in groups, further supporting the notion that echo chambers reinforce partisan divides by promoting homogeneous viewpoints. Similarly, findings from \textcite{cinelli2021echo} show that users tend to aggregate in homophilic groups on platforms such as Facebook and Twitter, around shared narratives. 

While substantial evidence supports the validity of selective exposure, which indicates that individuals often seek and disseminate information that confirms their pre-existing beliefs, the evidence regarding the avoidance of counter-attitudinal content remains inconclusive. In practice, people frequently encounter opposing viewpoints and may even engage with them \parencite{figa2022through,dubois2018echo,de2021no,vaccari2016echo,barbera2015tweeting,colleoni2014echo}. Moreover, the desire to seek pro-attitudinal content varies not only based on partisan alignments, but also according to the type of news consumed, political or apolitical \parencite{colleoni2014echo,barbera2015tweeting}. Research has indicated that online audiences may be less ideologically segregated compared to those utilizing offline channels, as they encounter cross-cutting counter-attitudinal content more frequently on social media \parencite{muise2022quantifying}.

Some studies have reported that exposure to cross-cutting news can facilitate the development of tolerant attitudes \parencite{wojcieszak2020can,chen2022effect}; however, other scholars caution that such exposure may backfire and exacerbate polarization -- ideological, issue and affective -- due to confirmation bias and motivated reasoning \parencite{bail2018exposure,guess2020does,kim2019cross,lin2023effects}. This underscores the notion that exposure to opposing viewpoints is sometimes ineffective in mitigating affective polarization \parencite{zhu2024implications}. Furthermore, evidence suggests that the polarizing effects of media exposure depend not only on the content (whether pro- or counter-attitudinal) but also on the dose of media exposure \parencite{arendt2015toward,lin2025exploring}.

\subsection{Elites and Affective Polarization}

Affective polarization is fundamentally rooted in the human propensity to categorize themselves into social groups, where individuals derive significant portions of their identity from these affiliations \parencite{iyengar2018strengthening}. This categorization fosters in-group favoritism and out-group discrimination, as theorized by \textcite{turner1979social}. In addition to the influence of social media platforms, political elites \textendash\  including politicians, media pundits, and opinion leaders \textendash\ further reinforce or alleviate these identities either by amplifying divisions or promoting consensus through their rhetoric and actions \parencite{skytte2021dimensions,huddy2021reducing,reiljan2024patterns}.

\textcite{banda2018elite} argue that the increasing levels of ideological extremity among the elites of opposing parties exacerbate affective polarization to a greater extent than the comparable levels of extremity among partisans’ own in-party elites. Moreover, they posit that this effect is amplified among individuals with higher levels of political interest. Similarly, \textcite{back2023elite} contend that polarized social identities are reinforced by partisan source cues, which shape perceptions of elite communication and contribute to heightened inter-group differentiation. In particular, the impact of such cues is more pronounced among voters with stronger partisan attachments.
Research also suggests that elite cooperation can mitigate affective polarization. \textcite{horne2023way}, for instance, find that inter-party collaboration in consensual political systems is associated with lower levels of inter-party hostility among the public, as supporters of governing parties express warmer sentiments toward coalition partners than can be explained by policy agreement alone. Likewise, \textcite{wagner2024elite} link the study of affective polarization with coalition politics, suggesting that when political elites signal a willingness to collaborate in coalition arrangements, they can reduce mutual animosity between partisan groups. Using an experimental approach, \parencite{huddy2021reducing} examine the effects of elite signaling on policy compromise and social interactions. Their findings indicate that narratives that emphasize warm personal relations between party leaders are particularly effective in alleviating affective polarization.

The models discussed thus far have empirically demonstrated the roles of social media, echo chambers, and political elites in exacerbating affective polarization. However, much of the existing research has examined these factors in isolation, leaving a critical gap in understanding how they interact and reinforce one another. For instance, while social media facilitates ideological sorting and cross-cutting exposure, its effects are contingent on elite rhetoric, which can either amplify partisan divides or foster consensus. Despite these interdependencies, the extent to which these forces jointly contribute to affective polarization remains under-explored. To address this gap, this study employs agent-based modeling (ABM) as a methodological approach that allows for the simulation of dynamic interactions among individuals, media environments, and political elites.

\subsection{Agent-based Models of Affective Polarization}

Agent-based methodologies, driven by recent advancements in computational capabilities, have gained widespread recognition in various disciplines, including economics, ecology, communication, and epidemiology \parencite{bonabeau2002agent,macy2002factors}. This approach involves modeling decentralized systems using dynamic, non-linearly interacting entities known as agents. These agents are characterized by specific attributes and governed by context-specific rules, creating artificial frameworks that simulate real-world systems. Such controlled environments enable researchers to focus on particular micro-level aspects of individual behavior and the underlying processes that drive them.

Although the origins of agent-based modeling can be traced back to the work of \textcite{neumann1966theory} on self-reproducing automata, its application within the social sciences is a relatively recent development. Notable early contributions to this field can be credited to \textcite{schelling1971dynamic}. However, it is in the past decade -- marked by significant advancements in computing capabilities -- that social scientists have increasingly embraced this methodology as a viable tool for understanding causal mechanisms \parencite{epstein1996growing,macy2002factors}. Agent-based models serve as intricate thought experiments that accommodate a wide range of inquiries, including social influence, structural differentiation, collective action, trust propagation, reputation dynamics, and homophily \parencite{macy2002factors}.

A fundamental question raised by \textcite{axelrod1997advancing} has ignited research aimed at understanding why individuals, despite a tendency to converge in beliefs, attitudes, and behaviors through interaction, do not become homogeneous. Early works by researchers such as \textcite{macy2003polarization} and \textcite{baldassarri2007dynamics} explored the dynamics of attitude formation and social polarization, considering factors such as ideological similarity and political interest in interpersonal interactions. In later work, \textcite{mas2013differentiation} examined the evolution of bipartisanship as a result of homophily, while \textcite{banisch2019opinion} delved into polarization within the context of opinion reinforcement among randomly communicating individuals situated in distinct homogeneous clusters. Recent research has expanded this discourse by exploring alternative explanations for bi-polar and multi-polar social systems, focusing on structural similarity, preferential heterogeneity, and elite-level reinforcement within social networks \parencite{santos2021link,tokita2021polarized,vasconcelos2021segregation,leonard2021nonlinear,macy2021polarization,martin2023multipolar}.

The agent-based models discussed thus far \parencite{macy2003polarization,baldassarri2007dynamics,mas2013differentiation,banisch2019opinion,santos2021link,tokita2021polarized,vasconcelos2021segregation,leonard2021nonlinear,macy2021polarization,martin2023multipolar} address various aspects of polarization, including issue-driven and ideological polarization among both elite individuals and the broader public. However, these models do not adequately capture the phenomenon of affective polarization, which is increasingly plaguing democracies worldwide -- not only posing a greater threat to democratic norms than issue-based or ideological polarization but also amplifying social and psychological divisions that extend beyond policy disagreements.

One of the first analytical models investigating the connection between exposure to diverse content on social media and the rise of affective polarization is presented in \textcite{tornberg2021modeling}. This model builds upon the reinforcement learning framework introduced by \textcite{banisch2019opinion}, wherein agents adapt their identities based on feedback from random pairwise interactions. The feedback received is positive when identities align and negative when they diverge, thereby accounting for exogenous variations in their social contexts.

Another relevant model presented by \textcite{tornberg2022digital} draws inspiration from \textcite{axelrod1997advancing}'s model of cultural dissemination. In this model, agents possess non-uniformly weighted ``cultures'' or identities, some of which are fixed. The agents are then arranged on a two-dimensional lattice and have the option to interact either locally with their neighbors or globally through randomized swapping. This design captures the impact of digital media, which often facilitates non-local interactions that extend beyond immediate social circles.

Additionally, \textcite{carpentras2023polarization} present an opinion dynamics model that integrates experimental data with a Social Identity Approach (SIA) to investigate affective shifts in individual attitudes and their impact on polarization. Notably, their model posits that attitudinal shifts resulting from interactions within an individual's in-group are more pronounced than those arising from interactions with out-group members.

While existing models \parencite{tornberg2022digital,tornberg2021modeling} primarily focus on explaining the rise of affective polarization as a consequence of information exchange through varied interpersonal interactions, they do not adequately account for a multitude of other factors that operate in parallel. These include partisan news consumption, the bias in its subsequent dissemination, political elites and the asymmetry in individual reactions to pro- and counter-attitudinal content. Therefore, we propose a model using an agent-based framework which investigates the interplay between three key drivers of affective polarization. The first is what we term as \textit{affective asymmetry}, or the differential manner in which partisans engage affectively with pro- and counter-attitudinal news content. The second is the network structure within which individual partisans are embedded in the media landscape. The third is the nature of the elite composition in the population.

Our model features a synthetic population characterized by partisan affiliations and affective attitudes toward both in-party and out-party groups. It challenges the traditional notion of the echo chamber by allowing agents embedded within the media landscape to encounter news content through two distinct channels. The first channel encompasses interactions with their network neighbors \textendash\ activities such as retweets, shares, or reposts \textendash\ while the second channel pertains to direct exposure to news content from original sources themselves. In this manner, our model allows agents to be exposed to information from their immediate neighbors (their ``echo-chamber'') as well as from news sources outside of those confines.

Additionally, our model incorporates diversity along party lines, within individual neighborhoods, mirroring real-world social interactions. Furthermore, it allows for asymmetrical affective engagement with news content through a gain function, reflecting the complex ways through which individuals interact with and respond to out-party versus in-party information. Moreover, we investigate the potential for polarizing effects to percolate from network elites to individuals with lesser influence, capturing the dynamics of opinion leaders in the polarization process. These features of our proposed model collectively provide a comprehensive framework for addressing key research questions on the mechanisms driving affective polarization in digital media environments.

\section{Methodology}
\subsection{Synthetic Population}

In our study, we investigate a synthetic group comprising a fixed population of N agents in the context of a two-party political system. Thus, each agent is affiliated with one of two parties, one of which we label the ``left-leaning party'' ($s_{1}$) and the other, ``the right-leaning party'' ($s_{2}$). 

We posit that an agent's partisan affiliation ($x_{i} \in \{s_{1},s_{2}\}$) plays a crucial role in shaping their emotional responses to information encountered on online platforms \parencite{iyengar2012affect}. Agents develop a sense of loyalty toward their respective party, termed In-Party Affect ($IPA_{i} \in IPA$), reflecting their positive sentiments and attachment to their own party. Conversely, they also experience hostility toward the opposing party, referred to as Out-Party Affect ($OPA_{i} \in OPA$), reflecting negative sentiments and animosity toward the opposing party.

Additionally, agents in the group, beyond their individual partisan affiliations and attitudes, are assumed to establish interpersonal relationships with other members. These social connections collectively constitute the underlying inter-personal network that we describe next.

\subsection{Social Network}

We utilize a directed and static scale-free network model, $G=(V,E)$, to depict the agents and their relationships within our synthetic group. This directed social network is a random graph in which both in-degree and out-degree of nodes follow power-law distributions with predefined exponents \parencite{goh2001universal,chung2002connected,cho2009percolation}. In this representation, network nodes correspond to individual agents within the group, while edges represent pathways for news propagation. Notably, these pathways are directional: Incoming edges signify potential news exposure, whereas outgoing edges indicate the dissemination of news to others. It is worth highlighting that the network's edges remain constant throughout our study, with no acquisitions or losses.

Interactions between agents result in the formation of distinct neighborhoods. An agent $i$'s neighborhood, denoted as $N_{i}$, comprises agents connected to agent $i$ through incoming edges that function as pathways for informational exposure for agent $i$.

Agents may possess varying in-degrees ($deg_{i}$), representing the number of connections within their neighborhood. Some agents may have a higher in-degree, indicating more connections, while others may have a lower in-degree with fewer connections. 

The partisan composition of the agents' neighborhoods varies greatly, with some agents surrounded predominantly by like-minded individuals (as in an echo chamber), while others are embedded in more balanced or even ideologically contrasting clusters. The ideological alignment within an agent $i$'s neighbors is quantified using a neighborhood similarity parameter (e.g., $p_{s_{i}}$), defined as:
\begin{equation}
p_{s_{i}} := \frac{|N_{s_{i}}|}{deg_{i}}, \forall i \in V,
\end{equation}
where, $|N_{s_{i}}|$ represents the count of neighbors who share the same ideology as Agent $i$. The parameter $p_{s_{i}}$ quantifies ideological similarity, ranging from complete congruence ($p_{s_{i}}=1$) to complete contrast ($p_{s_{i}}=0$). A value of $p_{s_{i}}=1$ indicates that agent $i$'s neighbors share identical partisan affiliations, while $p_{s_{i}}=0$ signifies diametrically opposing affiliations.

\subsection{Mechanisms of News Exposure and Spread}

In our model, we propose that agents within the group are exposed to news content through two distinct channels. The first involves their neighbors' activities, including actions like retweets, shares, or re-posts. The second involves direct exposure to news content from the original sources themselves. This captures the incidental aspect of information exposure which often occurs outside the agent's control. 

We assume that exposure to news content from sources is a probabilistic event, characterized by an exposure rate ($p_{e}$). This rate quantifies the likelihood of an agent encountering content from the news sources within the social network.

\begin{figure}[htp]
\centering
\includegraphics[width=\textwidth]{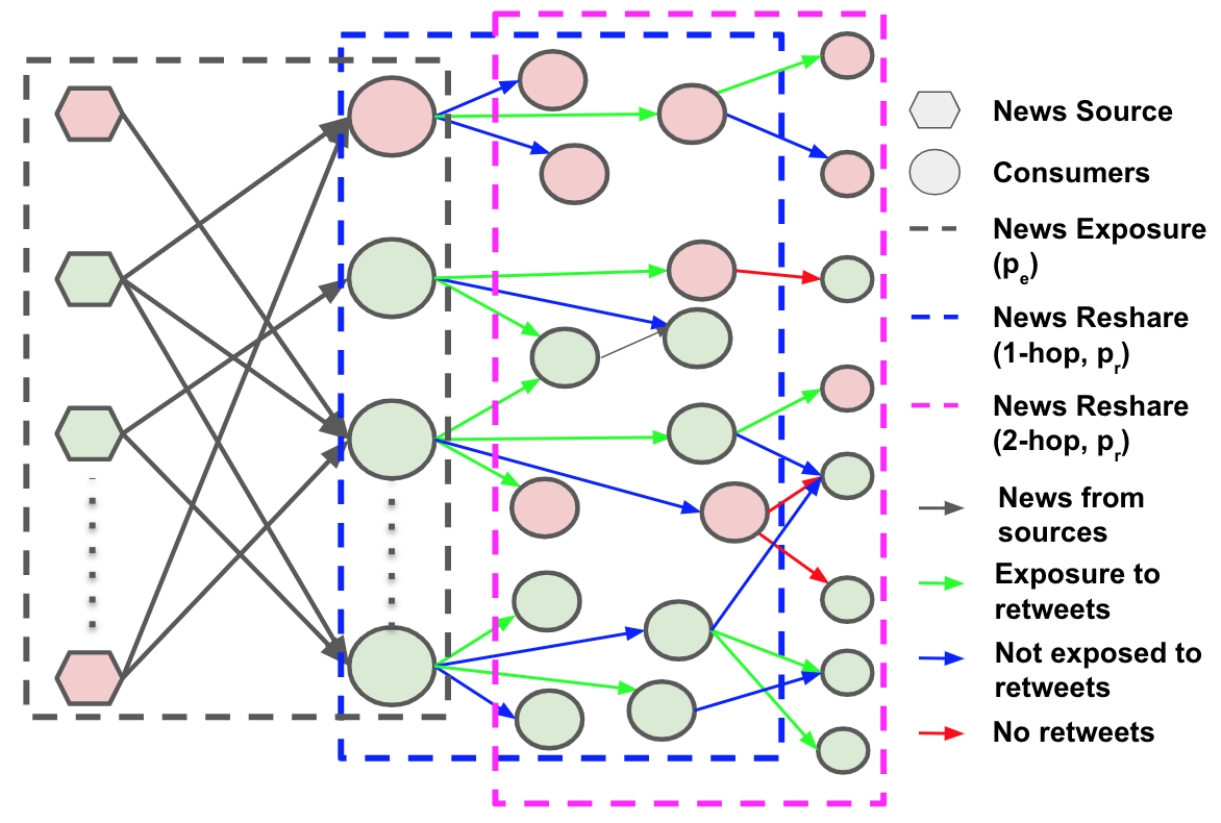}
\caption{Typical news spread mechanism in a social network\\
Note: The figure illustrates the spread of news in a social network driven by both media outlets and exposure to individual retweets. The process is assumed to be influenced by individual partisan affiliations. The light red and light green colors suggest the two predominant ideological slants.}
\label{fig1}
\end{figure}

Upon being exposed to this content, an agent can choose to share it with their neighbors. However, the dissemination of the content depends on a specific condition: agent $i$ will only share the news content with its neighbors, with a re-share probability $p_{r}$, if the ideological slant of the news aligns with their own \parencite{kalogeropoulos2017shares,wischnewski2021shareworthiness}. This alignment ensures that the agent is more inclined to propagate news that aligns with their existing beliefs or perspectives. We validate this assumption by means of an online survey, as shown in the Appendix B (see Results). Subsequently, the neighbors can choose to perpetuate the news cascade in subsequent iterations with the same re-share probability. This proposed mechanism of news exposure and subsequent spread among agents with matching slants is illustrated in Figure \ref{fig1}.

\subsection{Affective Response}
An individual's sentiments of loyalty and hostility are shaped by news content disseminated by media sources and reinforced through interpersonal connections. Exposure to news content that aligns with one's own ideological slant may strengthen their existing partisan sentiments and reinforce their loyalty towards their own party \parencite{iyengar2018strengthening,hasell2016partisan}. Conversely, exposure to news content with differing ideological slants may elicit feelings of animosity towards the opposing party \parencite{renstrom2023threats,groenendyk2018competing,iyengar2018strengthening,knobloch2020confirmation}. This assumption is validated by means of an online survey as detailed in the Appendix B (see Results).

The extent to which positive and negative sentiments are reinforced can vary widely among individuals. Some individuals may approach news content with a balanced perspective, adjusting their positive or negative sentiments to the same extent. On the other hand, some individuals may exhibit a strong bias in their reactions, favoring news content that confirms their pre-existing beliefs while disregarding information that challenges their worldview.

\begin{figure}[htp]
\centering
\includegraphics[width=\textwidth]{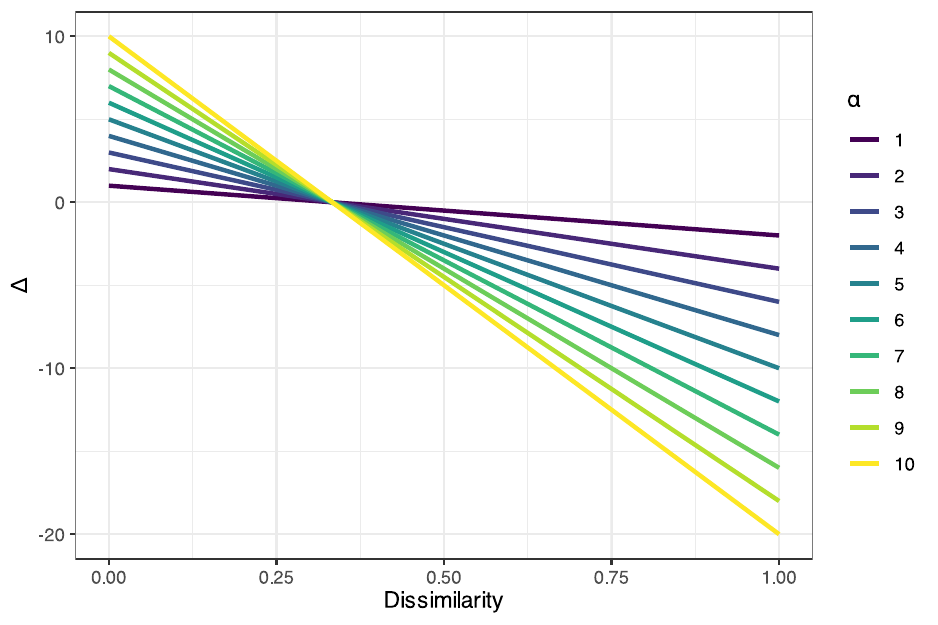}
\caption{Individuals' affective response to news content by slant dissimilarity and affective asymmetry\\
Note: The figure illustrates individuals' affective response as a function of slant dissimilarity upon exposure to news content, shown for varying levels of affective asymmetry ($\alpha$). Higher affective asymmetry ($\alpha$) results in a more pronounced affective response from individuals.}
\label{fig2}
\end{figure}

In our model, we aim to capture this range of individual responses to news exposure. We go beyond simply quantifying the effect of news exposure on individual partisan feelings based on the degree of dissimilarity in slants. We also consider the varying degrees of asymmetry in individuals' responses. Specifically, we define the individual response ($\Delta_{i}(t)$) upon exposure to news content, which is determined by the dissimilarity between an individual's partisan orientation and the slant of the news content they are exposed to ($s_{n_{i}}(t)$). This response is calculated using the following equation:
\begin{equation}
\Delta_{i}(t)~:=~ -3\times\alpha\times|x_{i}-s_{n_{i}}(t)| + \alpha.
\end{equation}
Here, the constant $\alpha\geq 1$ is used to adjust the asymmetry of individual responses to pro- and counter-attitudinal news content. The impact of any news exposure on an individual's feelings toward or against a particular party is determined using the following equations:
\begin{align}
IPA_{i}(t+1)~&:=~IPA_{i}(t)~+~\Delta_{i}(t),~\text{and} \label{e1}\\
OPA_{i}(t+1)~&:=~OPA_{i}(t)~+~\Delta_{i}(t).\label{e2}
\end{align}
When $\alpha = 1$, the magnitude of the individual response towards out-party content is twice as strong as the magnitude of response towards in-party content (but in the other direction). As depicted in Figure \ref{fig2}, although this trend can be observed for higher values of $\alpha$ as well, it's important to note that the asymmetry of individual response to both pro- and counter-attitudinal content is clearly exacerbated. For instance, when $\alpha = 10$, the response to pro-attitudinal content is ten times more intense, while the response to counter-attitudinal content is twenty times more extreme, compared to the case where $\alpha = 1$. 

Additionally, we quantify the emotional distance between an individual's in-party and out-party feelings using an affective distance, which serves as a measure of affective polarization at the individual level ($AD_{i}$). The measure of group affective polarization (AP) is defined as the mean affective distance across the entire population, providing an aggregate indicator of inter-party hostility prevailing within the group. We use a feeling thermometer scale that spans from 0 to 100 to quantify affective attitudes. Feeling thermometers are widely used in surveys to assess affective polarization.  Here, individuals indicate their feelings towards parties and politicians on a scale of 0 to 100 with values lower than 50 indicating feelings of hostility towards a party and values above that indicating feelings of warmth. \parencite{iyengar2019origins}.

\begin{align}
AP(t)~&:=\frac{\sum\limits_{i=1}^{N}AD_{i}(t)}{N} \label{e3},~\text{where},\\
AD_{i}(t+1)~&:=~IPA_{i}(t+1)~-~OPA_{i}(t+1).\label{e4}
\end{align}

\section{Simulations}
\subsection{Framework}
In this study, we simulate a synthetic cohort of $N=10,000$ individuals, each classified as either Left-leaning ($x_{i}=1$) or Right-leaning ($x_{i}=2$).  We maintain that these affiliations remain fixed throughout the simulation. 

The composition of this synthetic group, in terms of partisan affiliations, is determined by a population bias parameter, $p_{b} \in [0,1]$. When $p_{b}=0$ or $p_{b}=1$, the group is homogeneous, with all individuals uniformly aligned as either Left-leaning ($x_{i}=1,~\forall i\in V$) or Right-leaning ($x_{i}=2,~\forall i\in V$), respectively. Intermediate values of $p_{b}$, specifically $p_{b}=0.25$, $p_{b}=0.5$, and $p_{b}=0.75$, correspond to groups characterized by a Left-leaning majority, a balanced composition, and a Right-leaning majority, respectively.

The individuals in the synthetic population are assumed to be arranged on a static and directed social network, denoted as $G(V,E)$. This network comprises $N$ vertices ($|V|=N$) and $10,00,000$ directed edges ($|E|=10,00,000$), and is generated using the $igraph$ package in R \parencite{igraph}. In this network, individuals with the highest in-degrees, specifically those above the $90^{th}$ percentile ($p90$), are considered elite members ($V_{e}\subset V$) \textendash\ or the opinion leaders. We introduce the concept of an elite bias to configure the partisan composition of this elite group, allowing us to create scenarios with a Left-leaning majority ($p_{eb}=0.25$), a Right-leaning majority ($p_{eb}=0.75$), or a balanced composition ($p_{eb}=0.5$) among the elite opinion leaders. Incorporating this notion of an elite bias, distinct from the overall population bias, introduces a novel perspective into our comprehension of affective polarization, that has hitherto been overlooked in the literature. 

Both elite and non-elite members of the synthetic population are exposed to news content with an exposure rate, $p_{e}=0.01$. Each piece of news is explicitly partisan, either left- or right-leaning, but the distribution of news is uniform, so the probability of each message being left- or right-leaning is 0.5. When an individual is exposed to news content that aligns with their respective slant, we posit that they will invariably retweet it, with a reshare probability, $p_{r}=1$. However, the probability of these retweets propagating to their neighbors ($j\in N_{i}$) is determined by a retweet exposure rate, $p_{re}=0.5$. Although setting $p_{r}=1$ may appear to be a simplifying assumption, it does not qualitatively affect the final results of our model. This is because the likelihood of exposure to retweets -- which is more consequential in practice -- depends on the joint probability of both the reshare probability and the retweet exposure. All the parameters of the model and their respective values are summarised in Table \ref{table1}. 

\begin{table}[H]
\centering
  \begin{tabular}{@{}ccc@{}}         \toprule
  \multicolumn{3}{c}{Model Parameters}        \\ \cmidrule(r){1-3}
  Symbol    & Description & Value \\ \midrule
  $N$      & Total population    & $10,000$ \\
  $t_{f}$ & Total simulation time & $600$ \\
  $M$ & Total MonteCarlo iterations & $10$ \\
  $\alpha$ & Affective asymmetry & $\{1,5,10\}$ \\
  $s$          & Individual partisan affiliation        &  $\{1,2\}$ \\
  $p_{b}$       & Population bias     & $[0,1]$ \\
  $p_{eb}$       & Elite bias     & $[0,1]$ \\
  $IPA$       & Individual In-party affect     & $[50,100)$ \\
  $OPA$ & Individual Out-party affect    &  $(0,50)$ \\ 
  $G$ & Directed social network & $(V,E)$ \\
  $V$ & Vertex set & $\{1,2,\hdots,N\}$\\
  $E$ & Edge set & $\{(i,j): i,j \in V, i \neq j\}$\\
  $p_{e}$ & News exposure rate & $0.01$ \\
  $p_{r}$ & Retweet rate & $1$ \\
  $p_{re}$ & Retweet exposure rate & $0.5$ \\
  \bottomrule
  \end{tabular}
  \caption{Model parameters and corresponding values\\
Note: The table lists the various parameters of the proposed model along with their corresponding values.}
  \label{table1}
\end{table}

\subsection{Analyses}

The differences in composition between the general population and the elite subgroup result in qualitatively distinct network configurations. These networks allow for the examination of news consumption and affective reinforcement based on individuals' ideological alignment, in a range of network conditions characterized by various combinations of in-group majority/minority and in-group elite majority/minority. To facilitate this analysis, we simulate a total of nine distinct network configurations, as summarized in Table \ref{table2} and elaborated upon in the subsequent section. The simulations were conducted in R, and the source code has been deposited in a GitHub repository to facilitate replication:[ the link has been redacted for the sake of anonymity].

In our analysis, we simulate news consumption and sharing patterns within each of these nine synthetic social network configurations, spanning a duration of $t_{f}=600$ time units. These simulations adhere to the predefined parameters detailed in Table \ref{table1}. Since affective asymmetry -- the tendency for agents to react more strongly to out-party content they dislike than to in-party content they approve of -- is key to our model, we run simulations for each configuration using three different values of the asymmetry parameter: $\alpha=1$, $\alpha=5$, and $\alpha=10$. This allows us to examine how varying levels of asymmetry in affective response influence polarization dynamics. For each of the nine configurations, along with its specified affective asymmetry, we conduct repeated simulations ($M=10$), incorporating randomized initial values for individual slant, network structure (individual degree and neighborhood), and affective states.

From the simulation outcomes, we calculate the mean values for Affective Polarization, In-party Affect, and Out-party Affect. Furthermore, we assess the Time-to-Max-Affect in each scenario to compare the rate at which polarization evolves under the various conditions. Subsequent sections will detail the methodologies used to compute these values.

\begin{table}[H]
  \begin{tabular}{|c|c|c|c|}
\cline{1-4}
\multicolumn{1}{|c|}{\diagbox[trim=l,height=2\line]{\textbf{Elite bias}}{\textbf{Population bias}}} & \textbf{Left Majority} & \textbf{Balanced} & \textbf{Right Majority} \\
\cline{1-4}
\multicolumn{1}{|c|}{\textbf{Left Majority}} & Configuration-1 & Configuration-2 & Configuration-3 \\
\cline{1-4}
\multicolumn{1}{|c|}{\textbf{Balanced}} & Configuration-4 & Configuration-5 & Configuration-6 \\
\cline{1-4}
\multicolumn{1}{|c|}{\textbf{Right Majority}} & Configuration-7 & Configuration-8 & Configuration-9  \\
\cline{1-4}
\end{tabular}
\caption{Network configurations by population and elite subgroup compositions\\
Note: This is a tabulation of all possible configurations of social networks generated by different combinations of the compositions of general population and elite subgroups.}
\label{table2}
\end{table}

\subsubsection{Mean Affective Polarization}

This metric represents the average affective polarization observed within a particular network configuration over the course of $M$ Monte Carlo iterations, where initial values are randomized.
\begin{equation}
    \text{Mean Affective Polarization}(t)~:=~\frac{\sum\limits_{i=1}^{M} \text{AP}_{i}(t)}{M},~\forall t \leq t_{f}
\end{equation}

\subsubsection{Mean In-Party Affect}

This metric signifies the average in-party affect, which reflects the ``warm'' sentiments of loyalty within the synthetic population. It is observed within a specific network configuration across $M$ Monte Carlo iterations, with initial values randomized.
\begin{align}
    \text{Mean In-party Affect}(t)~&:=~\frac{\sum\limits_{i=1}^{M} (\text{In-party Affect}(t))_{i}}{M},~\forall t \leq t_{f},~\text{where},\\
    \text{In-party Affect}(t)~&:=~\frac{\sum\limits_{j=1}^{N} \text{IPA}_{j}(t)}{N}.
\end{align}

\subsubsection{Mean Out-Party Affect}

It represents the average out-party affect, indicating the ``cold'' feelings of hostility within the synthetic population, observed in a specific network configuration across $M$ Monte Carlo iterations with randomized initial values.
\begin{align}
    \text{Mean Out-party Affect}(t)~&:=~\frac{\sum\limits_{i=1}^{M} (\text{Out-party Affect}(t))_{i}}{M},~\forall t \leq t_{f},~\text{where},\\
    \text{Out-party Affect}(t)~&:=~\frac{\sum\limits_{j=1}^{N} \text{OPA}_{j}(t)}{N}.
\end{align}

\subsubsection{Mean Time-to-Max-Affect}

This metric represents the average time required to reach maximum affective polarization within a specific network configuration across $M$ Monte Carlo iterations, with initial values randomized. \begin{equation}
t_{\text{max}} = \frac{\sum\limits_{i=1}^{M} (t_{90})_{i}}{M}.
\end{equation}
Here, $t_{90}$ refers to the duration it takes for affective polarization to reach 90\% of its maximum value in each individual simulation run.

\section{Results}

\subsection{Configurations 1, 5 and 9: Matching Ideological Composition Across Groups}

Configurations 1, 5 and 9 represent scenarios where the ideological composition of both the elite subgroup and the larger population is identical. In Configuration-1 (or Configuration-9), a significant majority of members within both the general population and the elite subgroup exhibit an affiliation towards the Left-leaning (or Right-leaning) party. This leads to neighborhoods that are predominantly composed of individuals with similar ideological alignments ($p_{s_{\text{mean}}}\approx 0.75$). Consequently, when individuals within these neighborhoods share news content that aligns with their own slant, it is widely consumed and further disseminated. This triggers a significant increase in loyalty to one's own party as evidenced by an approximately 50\% surge in their in-party affect within a time frame of 100 units, as illustrated in Figure \ref{fig4}.

Conversely, in Configuration-1 and Configuration-9, where ideologically dissimilar neighbors form a minority, they respond negatively to these retweets. This adverse reaction results in nearly a 25\% decrease in their out-party affect within 100 units of time (see Figure \ref{fig5}). Although the concurrent increase in in-party affect among the majority and the decrease in out-party affect among the minority are comparable at the end of the simulation period, the majoritarian phenomenon predominantly drives the escalation of affective polarization (see Figure \ref{fig3}). The asymmetry in individuals' affective responses amplifies this phenomenon, resulting in a 35\% increase when $\alpha=5$ and a 44\% increase when $\alpha=10$, in comparison to when $\alpha=1$ as illustrated in Figure \ref{fig3}. Moreover, Figure \ref{fig6} reveals a notable reduction in Time-to-Max-Affect by almost 45\% when $\alpha$ transitions from 5 to 10.  

In a group where the number of individuals from both partisan affiliations is approximately equal (Configuration-5), the composition of neighborhoods differs from Configuration-1 and Configuration-9. Here, there is a reduced presence of like-minded neighbors and a corresponding increase in dissimilar neighbors ($p_{s_{\text{mean}}}\approx 0.5$). Consequently, each retweet of news content not only fosters increased feelings of loyalty among like-minded members but also triggers a significant rise in hostility among dissimilar neighbors. This is evident in the nearly 42\% increase in in-party affect (see Figure \ref{fig4}) and the 87.5\% decrease in out-party affect (see Figure \ref{fig5}) within 100 units of time. Consequently, the affective polarization in the group rises to nearly 80\% during the same time period (see Figure \ref{fig3}).
The substantial reduction in out-party affect relative to the increase in in-party affect in a balanced population is primarily driven by the affective asymmetry in individual responses. Moreover, varying the affective asymmetry from 1 to 10 accelerates the mean Time-to-Max-Affect by almost 97\% (see Figure \ref{fig6}).

\subsection{Configurations 3 and 7: Ideological Opposition Between Elite and Population}

In Configuration-3 (or Configuration-7), the elite subgroup predominantly exhibits a Left-leaning majority (or Right-leaning majority), while the larger population holds an opposing majority stance, primarily being Right-leaning (or Left-leaning). Consequently, elite neighborhoods comprise ideologically opposed individuals, with an average similarity index of approximately $p_{s_{\text{mean}}}\approx 0.25$, whereas like-minded individuals constitute the neighborhoods in the remainder of the population, with an average similarity index of approximately $p_{s_{\text{mean}}}\approx 0.75$.

Consequently, the affective polarization within the synthetic group primarily stems from an increase in out-party affect within elite neighborhoods and a rise in in-party affect within neighborhoods comprising the remainder of the population. Within a time frame of 100 units, there is a nearly 37.5\% decrease in out-party affect (see Figure \ref{fig4}) and an almost 50\% increase in in-party affect (see Figure \ref{fig5}). While the rise in In-party affect is akin to that observed in Configuration-1 and Configuration-9, there is an additional decline of 12.5\% in out-party affect due to the presence of ideologically opposed neighborhoods among elites. In line with prior findings, an increase in affective asymmetry from $\alpha=5$ to $\alpha=10$ not only augments affective polarization by 28\% but also reduces the mean Time-to-Max-Affect by nearly 40\% (see Figures \ref{fig3} and \ref{fig6}).

\subsection{Configurations 2 and 8: Biased Elites amidst a Balanced Public}

In Configuration-2 (or Configuration-8), the elite subgroup predominantly leans Left (or Right), while the larger population exhibits a balanced composition with no clear majority. Consequently, both elite and non-elite neighborhoods consist of individuals with Left-leaning and Right-leaning inclinations, resulting in an average similarity index of approximately 0.5.

Within the neighborhoods of the larger population, the consumption and retweeting of news content with a similar slant lead to a proportional increase in in-party affect and a decrease in out-party affect. Although this phenomenon bears similarities to that observed in the neighborhoods of Configuration-5, the resulting increase in in-party affect (see Figure \ref{fig4}) at the end of the simulation period is nearly 20\% lower in comparison. Similarly, the decrease in out-party affect is observed to be 10\% lower (see Figure \ref{fig5}) than in Configuration-5. This divergence can be attributed to the presence of a polarized elite subgroup that resists the consumption and subsequent dissemination of counter-attitudinal news, in contrast to the more balanced composition of the rest of the population. Nonetheless, the influence of affective asymmetry in exacerbating affective polarization within the overall group is evident from the trends in Figures \ref{fig3} and \ref{fig5}.

\subsection{Configurations 4 and 6: Balanced Elites amidst a Biased Public}

In Configuration-4 (or Configuration-6), the elite subgroup is balanced, while the rest of the population has a Left-leaning (or a Right-leaning) majority. Consequently, although the neighborhoods of both elites and the rest of the population are composed mostly of Left-leaning (or Right-leaning) individuals, the average values of their similarity indices are approximately 0.5 and 0.75, respectively.

\begin{figure}[htp]
\centering
\includegraphics[width=\textwidth]{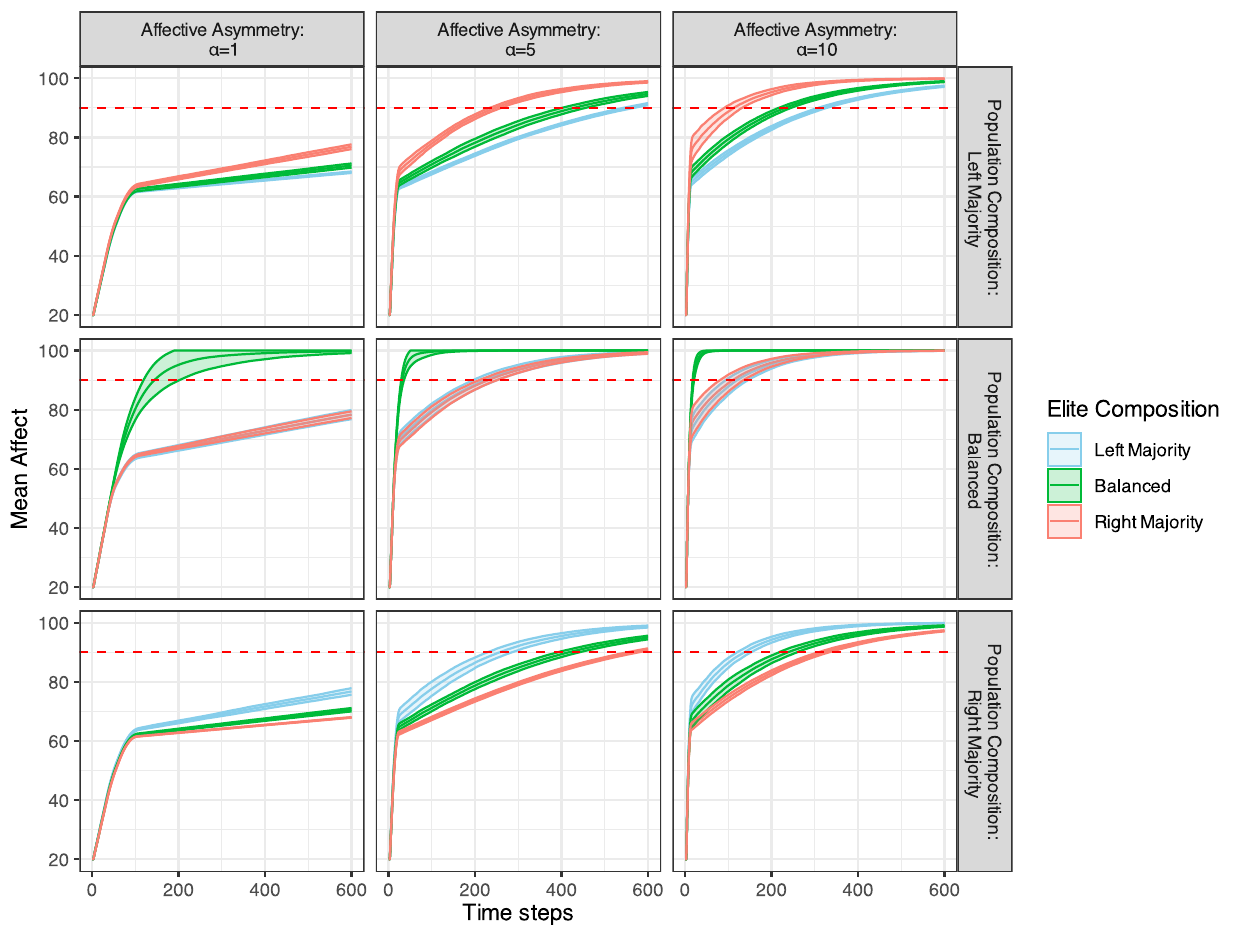}
\caption{Variation in mean affective polarization by affective asymmetry, population composition, and elite subgroup composition\\
Note: The figure illustrates the variation in mean affective polarization as a function of affective asymmetry, population composition, and elite subgroup composition. The solid line in the center of the line plot denotes the mean value, while the shaded region of the same color represents the standard deviation. The increase in mean affect is most rapid when the composition of both the elite subgroup and the general population is balanced.}
\label{fig3}
\end{figure}

\begin{figure}[htp]
\centering
\includegraphics[width=\textwidth]{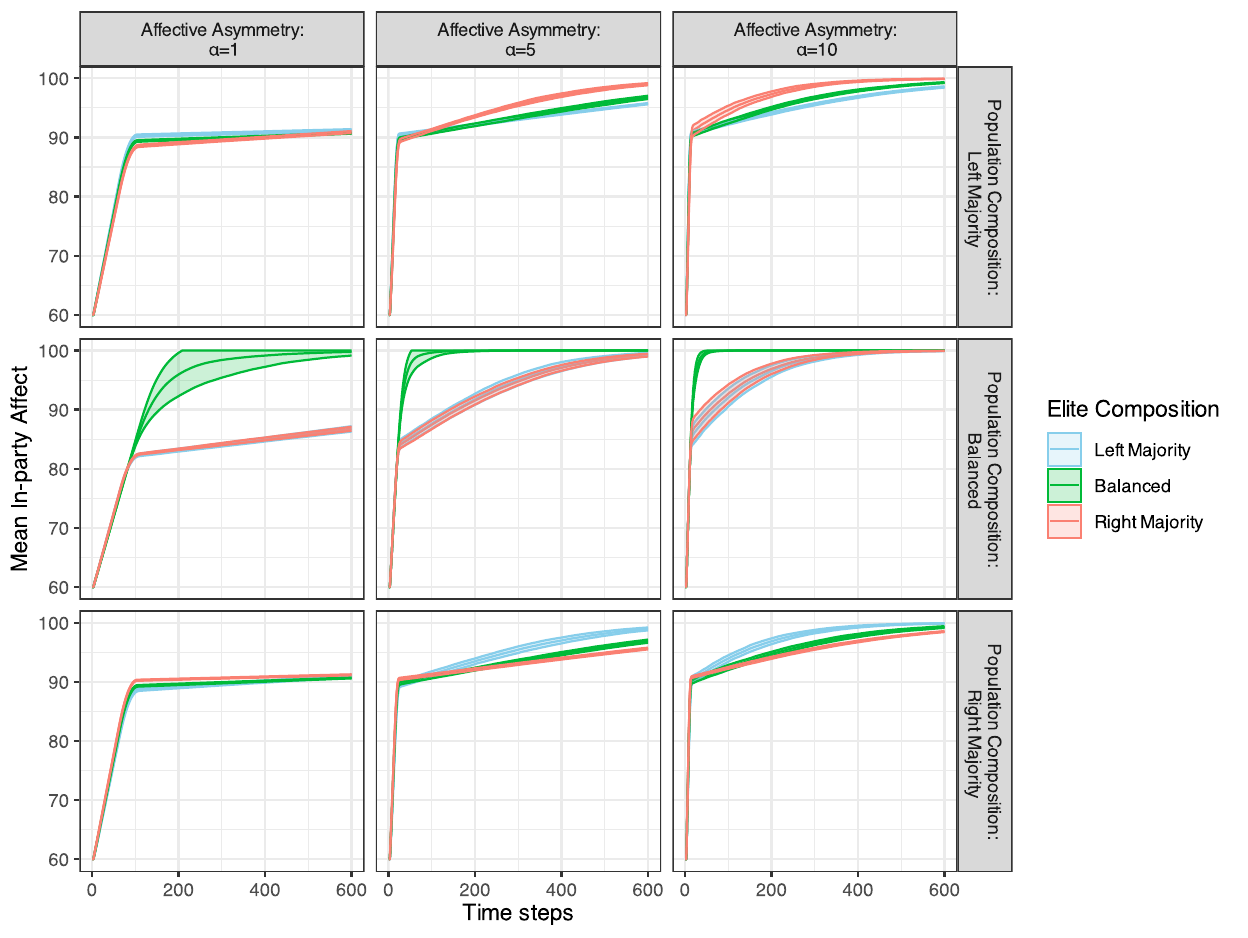}
\caption{Variation in mean in-party affect by affective asymmetry, population composition, and elite subgroup composition\\
Note: The figure illustrates the variation in mean in-party affect as a function of affective asymmetry, population composition, and elite subgroup composition. The solid line in the center of the line plot denotes the mean value, while the shaded region of the same color represents the standard deviation. The increase in mean in-party affect is most rapid when the composition of both the elite subgroup and the general population is balanced.}
\label{fig4}
\end{figure}

\begin{figure}[htp]
\centering
\includegraphics[width=\textwidth]{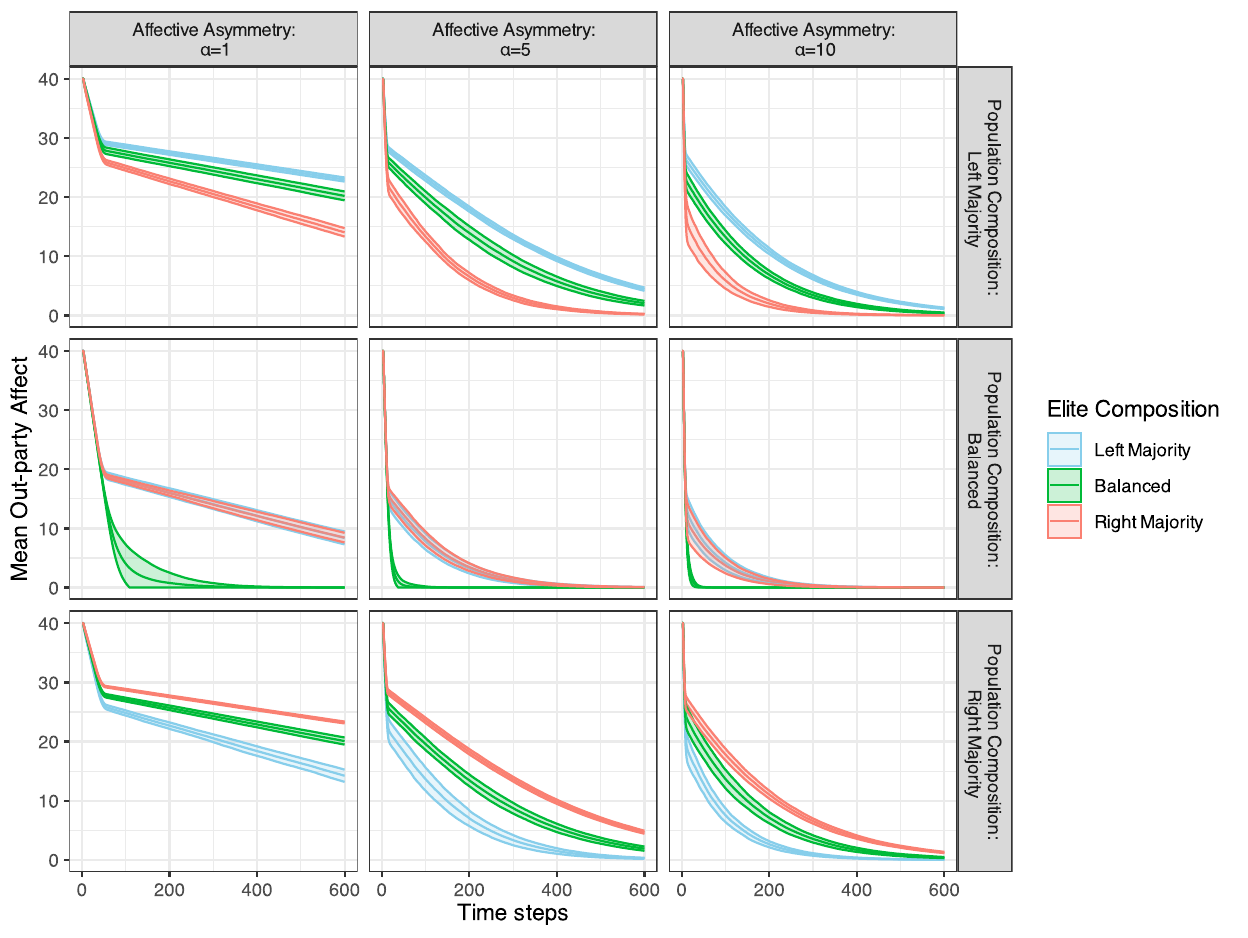}
\caption{Variation in mean out-party affect by affective asymmetry, population composition, and elite subgroup composition\\
Note: The figure illustrates the variation in mean out-party affect as a function of affective asymmetry, population composition, and elite subgroup composition. The solid line in the center of the line plot denotes the mean value, while the shaded region of the same color represents the standard deviation. The decrease in mean out-party affect is most rapid when the composition of both the elite subgroup and the general population is balanced.}
\label{fig5}
\end{figure}

Members of the larger population have like-minded neighborhoods, so when a news content is retweeted, it gets consumed and propagated by the like-minded majority, thereby increasing the in-party affect (see Figure \ref{fig4}). However, it also generates an adverse response among the dissimilar minority in the neighborhood, resulting in a corresponding decrease in the out-party affect (see Figure \ref{fig5}). However, the cumulative increase in affective polarization is predominantly determined by the like-minded majority neighbors (see Figure \ref{fig3}). This is similar to our observations in Configuration-1 (or Configuration-9) where both the population and the elite subgroup had a Left-leaning (or Right-leaning) majority.

\begin{figure}[htp]
\centering
\includegraphics[width=\textwidth]{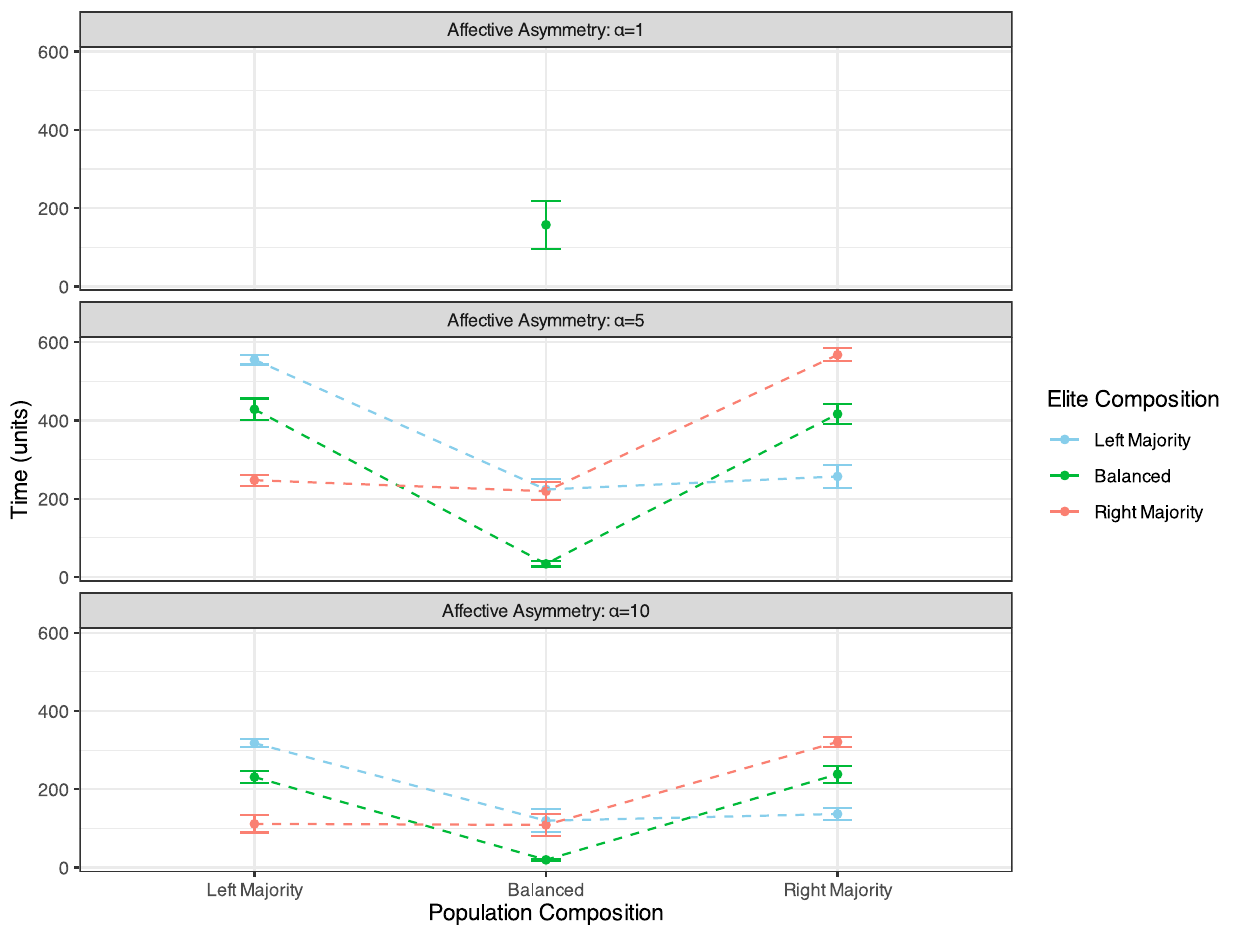}
\caption{Variation in mean time-to-max-affect by affective asymmetry, population composition, and elite subgroup composition\\
Note: The figure illustrates the variation in mean Time-to-Max-Affect ($t_{max}$) as a function of affective asymmetry, population composition, and elite subgroup composition. The point values denote mean time-to-max-affect, while the bars represent the standard deviation. Notably, for certain configurations with $\alpha=1$, some observations are missing since the maximum affect value was not attained.}
\label{fig6}
\end{figure}

On the contrary, the absence of a clear majority among the elite subgroup results in a substantial decrease in the out-party affect when the retweeted content does not match the ideological alignment of the Left-leaning (or Right-leaning) majoritarians in the neighborhood. Consequently, the out-party affect in Configuration-4 (or Configuration-6) decreases by 15\% more within 100 units of time and by 37.5\% (see Figure \ref{fig5}) more by the end of the simulation period, relative to Configuration-1 (or Configuration-9). Consistent with our observations in other configurations, an increase in affective asymmetry from $\alpha=5$ to $\alpha=10$, results in an increase in affective polarization by almost 14\%. This can also be observed from Figure \ref{fig6}, where the time taken to reach maximum affect is lowered by nearly 49\%.

Our simulations suggest that a population that is balanced (i.e. has roughly equal number of partisans on both sides of the spectrum), witnesses faster dissemination of news content representing diverse ideologies, consequently intensifying affective polarization quicker than majoritarian populations (i.e. where one group commands a majority). Our findings provide a comprehensive analysis of affective polarization stemming from news consumption and its dissemination under various network conditions and different levels of elite bias. This analysis encompasses three critical dimensions: (1) affective asymmetry, (2) the composition of individual neighborhoods, and (3) the composition of elite groups.

Conversely, when the broader public demonstrates ideological balance while the elite groups exhibit majoritarianism (i.e. there are more elites belonging to one group than the other), the rise in affective polarization is tempered due to the restricted consumption of counter-attitudinal content by the elites. Additionally, the more negatively individuals respond to out-party content relative to in-party content, more intense is this phenomenon. Taken in sum, our findings indicate that all three aspects of our model play a pivotal role in delineating the causal mechanism underlying the proliferation and escalation of affective polarization on digital platforms.

\section{Discussion}

This study presents a formal agent-based model to examine the mechanisms underlying affective polarization, characterized by heightened animosity between political partisans \parencite{iyengar2019origins}. The model simulates how the consumption and dissemination of ideologically skewed news content within digital media environments contribute to polarization. Synthetic agents are governed by parameters that define partisan affiliation, in-group and out-group affect, and affective asymmetry. Additional parameters regulate media dynamics, including news bias, frequency of exposure, and dissemination rates, allowing for systematic exploration of the effects of different media environments.

In light of growing empirical evidence challenging the conventional filter bubble and echo chamber hypotheses \parencite{figa2022through,dubois2018echo,de2021no,vaccari2016echo,barbera2015tweeting,colleoni2014echo}, our model incorporates incidental exposure to counter-attitudinal news content. While incidental exposure as a causal mechanism has been previously explored in agent-based models \parencite{tornberg2022digital,carpentras2023polarization}, the novelty of our approach lies in the users' agency to actively curate the information they perpetuate. Specifically, our model highlights how confirmation bias drives the widespread diffusion of news throughout the network, mimicking a key aspect of information cascade mechanisms observed in real-world social media platforms \parencite{barbera2015birds,garimella2018quantifying}.  Additionally,  we  also identify three key factors that may drive affective polarization: (1) affective asymmetry, (2) the proportion of partisan neighbors, and (3) elite group composition in the overall social network. Factors (2) and (3) are operationalized within 
network structures to generate nine distinct configurations, enabling a comprehensive evaluation of news consumption patterns and affective reinforcement. Consequently, unlike some agent-based models \parencite{tornberg2022digital,tornberg2021modeling} that primarily attribute polarization to interpersonal interactions, our approach highlights the role of structural factors, such as network composition -- alongside inter-personal interactions -- in shaping affective polarization. 

Our results also indicate that affective asymmetry -- where individuals exhibit stronger negative emotions toward out-groups than positive emotions toward in-groups -- plays a central role in determining how quickly affective 
polarization manifests at the population level. Additionally, network composition, particularly the proportion of partisan neighbors and elite group configurations, significantly influences the evolution of polarization. Contrary to studies suggesting that exposure to cross-cutting content fosters tolerance \parencite{chen2022effect,wojcieszak2020can}, our findings suggest that such exposure does not necessarily mitigate affective polarization, particularly in contexts where both elites and the general public are ideologically balanced. Instead, consistent with \textcite{lin2023effects}, cross-partisan exposure can exacerbate affective divides by amplifying animosity toward opposing groups. These findings challenge the assumption that increased ideological diversity in news consumption inherently reduces polarization, underscoring the need for further empirical research into conditions under which cross-cutting exposure mitigates or intensifies affective polarization.

Moreover, our simulations show that ideological balance -- where partisan groups are of roughly equal size -- accelerates the dissemination of ideologically diverse news, leading to faster intensification of affective polarization compared to majoritarian populations, where one group dominates. This aligns with \textcite{back2023elite}, who argue that polarized social identities are reinforced by partisan cues shaping perceptions of elite communication. However, when the general public is ideologically balanced while elites remain skewed toward one side (i.e., elite majoritarianism), the escalation of affective polarization is moderated. This suggests that elite bias has an outsize influence on the extent of counter-attitudinal content exposure, with more ideologically homogeneous elites limiting such exposure. Furthermore, polarization is more pronounced when individuals respond more negatively to out-party content than they do positively to in-party content. Collectively, these findings underscore the significance of (1) affective asymmetry, (2) partisan neighborhood composition, and (3) elite group composition in shaping affective polarization in digital spaces.

The most significant implication of our study is the need to reevaluate the normative assumptions surrounding affective polarization. As previously noted, our findings suggest that while an ideological balance within a group provides for more rapid growth in affective polarization, it simultaneously facilitates cross-cutting exposure to diverse viewpoints. Specifically, the presence of individuals with varying ideological affiliations within each neighborhood ensures the dissemination of news from multiple ideological perspectives, thereby broadening the range of narratives to which individuals are exposed. This outcome of our proposed model contributes to the ongoing discourse on the implications of polarization for democratic societies. 
Specifically, it provides empirical support for\textcite{kreiss2024review}'s provocation, which posits that polarization may not be inherently detrimental to democracy. Instead, it may be seen as an inevitable byproduct of democratic processes that encourage pluralism and debate. For instance, they argue that the Black Lives Matter movement was amplified by online platforms, allowing it to garner significant attention -- across the press, public and political spheres -- catalyzing action towards social justice. While its heightened visibility undoubtedly led to social and political polarization, it also increased support for the movement. Thus, \textcite{kreiss2024review} emphasize the importance of context in understanding polarization, noting that while it can lead to fragmentation, it can also promote vibrant democratic discourse when individuals are exposed to a variety of viewpoints. By highlighting the potential benefits of ideological diversity, our study aligns with \textcite{kreiss2024review}'s
assertion that polarization can serve as a  mechanism for fostering engagement with diverse perspectives. However, as their study further
asserts, it is crucial to recognise that platforms operate within political societies characterized by complex and often contentious histories regarding pluralism and justice. Therefore, future research must account for these historical and societal dynamics for a holistic examination of the platformed evolution of affective polarization.

Secondly, our findings on affective asymmetry are particularly consequential in current digital environments where outrage-driven engagement is amplified by algorithmic curation. Consistent with prior research \parencite{kalogeropoulos2017shares,wischnewski2021shareworthiness}, our survey results further validate our modeling assumption that individuals are more likely to share ideologically congruent news while reacting negatively to opposing viewpoints \parencite{renstrom2023threats,groenendyk2018competing,iyengar2018strengthening,knobloch2020confirmation}. Furthermore, exposure to counter-attitudinal content can provoke stronger negative reactions rather than fostering ideological openness, indicating that social media's role in polarization is more complex than previously assumed. Future research should explore whether platform-level interventions -- such as reducing the visibility of outrage-inducing content -- could mitigate these effects without undermining democratic discourse.

Elites also plays a pivotal role in enabling the growth of affective polarization. The composition of elite groups in digital networks influences ideological reinforcement. Political elites -- such as elected officials, media figures, and influencers -- serve as key nodes in disseminating partisan content. Our findings suggest that elite majoritarianism, where one ideological group dominates elite discourse while the public remains ideologically balanced, may temper the rise of affective polarization by restricting exposure to counter-attitudinal content. This highlights the intertwined nature of elite-driven and public polarization. Future studies can investigate whether interventions promoting bipartisan discourse or discouraging incendiary rhetoric could mitigate mass affective polarization.

As with any model, our approach simplifies real-world complexities and does not capture all aspects of digital media environments. Several avenues for refinement remain. First, news sources vary in credibility, and individuals assign different levels of trust to various outlets, as evidenced in our survey (see Results in Appendix B). Future models could integrate source-level trust to more closely reflect real-world media consumption;  When individuals trust partisan outlets, they are more likely to consume and share ideologically skewed content as evidenced in our survey (see Results in Appendix B). Second, while our model assumes that each piece of news is explicitly partisan -- either left- or right-leaning -- the distribution of news is uniform. However, real-world media landscapes are more complex and often contain inherent biases that play a crucial role in shaping public perception. Incorporating a more granular parameter, such as one that captures media bias, into future simulations could yield a more nuanced understanding of its role in polarization. Third, while our assumption that individuals predominantly retweet content aligned  with their ideologies reflects past empirical patterns \parencite{barbera2015birds,garimella2018quantifying}, it may oversimplify the complex dynamics observed in real-world systems. For instance, as several past studies have pointed out, this behavior is observed primarily in the case of identity-affirming topics and not in others \parencite{colleoni2014echo,barbera2015tweeting}. This suggests that the extent to which cross-cutting exposure exacerbates or mitigates affective polarization may be contingent on the salience of the issue to partisan identity, highlighting the need for a more nuanced understanding of polarization dynamics across different issue domains. Finally, the assumption of homogeneous neighborhood structures oversimplifies the complexity of social interactions, which future research should address.

Given these limitations, our findings should be interpreted within the model’s constraints and complemented with further empirical work to validate key mechanisms. Despite these caveats, our study provides a foundational framework for holistically understanding affective polarization in digital media environments. We encourage future research to refine and build upon this model to further investigate the interplay between media dynamics, elite discourse, and affective polarization.

\nocite{}
\printbibliography

@article{cinelli2021echo,
  title={The echo chamber effect on social media},
  author={Cinelli, Matteo and De Francisci Morales, Gianmarco and Galeazzi, Alessandro and Quattrociocchi, Walter and Starnini, Michele},
  journal={Proceedings of the National Academy of Sciences},
  volume={118},
  number={9},
  pages={e2023301118},
  year={2021},
  publisher={National Acad Sciences}
}

@article{ross2022echo,
  title={Echo chambers, filter bubbles, and polarisation: A literature review},
  author={Ross Arguedas, Amy and Robertson, C and Fletcher, Richard and Nielsen, R},
  year={2022},
  publisher={Reuters Institute for the Study of Journalism}
}

@article{levy2021social,
  title={Social media, news consumption, and polarization: Evidence from a field experiment},
  author={Levy, Ro’ee},
  journal={American economic review},
  volume={111},
  number={3},
  pages={831--870},
  year={2021},
  publisher={American Economic Association 2014 Broadway, Suite 305, Nashville, TN 37203}
}

@article{de2021no,
  title={No echo in the chambers of political interactions on Reddit},
  author={De Francisci Morales, Gianmarco and Monti, Corrado and Starnini, Michele},
  journal={Scientific reports},
  volume={11},
  number={1},
  pages={2818},
  year={2021},
  publisher={Nature Publishing Group UK London}
}

@article{muise2022quantifying,
  title={Quantifying partisan news diets in Web and TV audiences},
  author={Muise, Daniel and Hosseinmardi, Homa and Howland, Baird and Mobius, Markus and Rothschild, David and Watts, Duncan J},
  journal={Science advances},
  volume={8},
  number={28},
  pages={eabn0083},
  year={2022},
  publisher={American Association for the Advancement of Science}
}

@article{vaccari2016echo,
  title={Of echo chambers and contrarian clubs: Exposure to political disagreement among German and Italian users of Twitter},
  author={Vaccari, Cristian and Valeriani, Augusto and Barber{\'a}, Pablo and Jost, John T and Nagler, Jonathan and Tucker, Joshua A},
  journal={Social media+ society},
  volume={2},
  number={3},
  pages={2056305116664221},
  year={2016},
  publisher={Sage Publications Sage UK: London, England}
}

@article{colleoni2014echo,
  title={Echo chamber or public sphere? Predicting political orientation and measuring political homophily in Twitter using big data},
  author={Colleoni, Elanor and Rozza, Alessandro and Arvidsson, Adam},
  journal={Journal of communication},
  volume={64},
  number={2},
  pages={317--332},
  year={2014},
  publisher={Oxford University Press}
}

@article{renstrom2023threats,
  title={Threats, emotions, and affective polarization},
  author={Renstr{\"o}m, Emma A and B{\"a}ck, Hanna and Carroll, Royce},
  journal={Political Psychology},
  volume={44},
  number={6},
  pages={1337--1366},
  year={2023},
  publisher={Wiley Online Library}
}

@article{banda2018elite,
  title={Elite polarization, party extremity, and affective polarization},
  author={Banda, Kevin K and Cluverius, John},
  journal={Electoral Studies},
  volume={56},
  pages={90--101},
  year={2018},
  publisher={Elsevier}
}

@article{enders2021issues,
  title={Issues versus affect: How do elite and mass polarization compare?},
  author={Enders, Adam M},
  journal={The Journal of Politics},
  volume={83},
  number={4},
  pages={1872--1877},
  year={2021},
  publisher={The University of Chicago Press Chicago, IL}
}

@article{iyengar2018strengthening,
  title={The strengthening of partisan affect},
  author={Iyengar, Shanto and Krupenkin, Masha},
  journal={Political Psychology},
  volume={39},
  pages={201--218},
  year={2018},
  publisher={Wiley Online Library}
}

@article{groenendyk2018competing,
  title={Competing motives in a polarized electorate: Political responsiveness, identity defensiveness, and the rise of partisan antipathy},
  author={Groenendyk, Eric},
  journal={Political Psychology},
  volume={39},
  pages={159--171},
  year={2018},
  publisher={Wiley Online Library}
}

@article{santos2021link,
  title={Link recommendation algorithms and dynamics of polarization in online social networks},
  author={Santos, Fernando P and Lelkes, Yphtach and Levin, Simon A},
  journal={Proceedings of the National Academy of Sciences},
  volume={118},
  number={50},
  pages={e2102141118},
  year={2021},
  publisher={National Acad Sciences}
}

@article{jost2022cognitive,
  title={Cognitive--motivational mechanisms of political polarization in social-communicative contexts},
  author={Jost, John T and Baldassarri, Delia S and Druckman, James N},
  journal={Nature Reviews Psychology},
  volume={1},
  number={10},
  pages={560--576},
  year={2022},
  publisher={Nature Publishing Group US New York}
}

@article{gonzalez2023social,
  title={Do social media undermine social cohesion? A critical review},
  author={Gonz{\'a}lez-Bail{\'o}n, Sandra and Lelkes, Yphtach},
  journal={Social Issues and Policy Review},
  volume={17},
  number={1},
  pages={155--180},
  year={2023},
  publisher={Wiley Online Library}
}

@article{castle2021partisanship,
  title={Partisanship, religion, and issue polarization in the United States: A reassessment},
  author={Castle, Jeremiah J and Stepp, Kyla K},
  journal={Political Behavior},
  pages={1--25},
  year={2021},
  publisher={Springer}
}

@article{robison2019group,
  title={The group basis of partisan affective polarization},
  author={Robison, Joshua and Moskowitz, Rachel L},
  journal={The Journal of Politics},
  volume={81},
  number={3},
  pages={1075--1079},
  year={2019},
  publisher={The University of Chicago Press Chicago, IL}
}

@article{knudsen2021affective,
  title={Affective polarization in multiparty systems? Comparing affective polarization towards voters and parties in Norway and the United States},
  author={Knudsen, Erik},
  journal={Scandinavian Political Studies},
  volume={44},
  number={1},
  pages={34--44},
  year={2021},
  publisher={Wiley Online Library}
}

@article{neyazi2023political,
  title={Political Campaign Ads on Facebook: Investigating the Effects of Incivility in Videos and User Comments on Affective Polarization and Mobilization},
  author={Neyazi, Taberez Ahmed and Kuru, Ozan and Mukerjee, Subhayan},
  journal={International Journal of Communication},
  volume={17},
  pages={24},
  year={2023}
}

@article{jurkowitz2020us,
  title={US media polarization and the 2020 election: A nation divided},
  author={Jurkowitz, Mark and Mitchell, Amy and Shearer, Elisa and Walker, Mason},
  journal={Pew Research Center},
  volume={24},
  year={2020}
}

@article{hobolt2021divided,
  title={Divided by the vote: Affective polarization in the wake of the Brexit referendum},
  author={Hobolt, Sara B and Leeper, Thomas J and Tilley, James},
  journal={British Journal of Political Science},
  volume={51},
  number={4},
  pages={1476--1493},
  year={2021},
  publisher={Cambridge University Press}
}

@article{kekkonen2021affective,
  title={Affective blocs: Understanding affective polarization in multiparty systems},
  author={Kekkonen, Arto and Yl{\"a}-Anttila, Tuomas},
  journal={Electoral Studies},
  volume={72},
  pages={102367},
  year={2021},
  publisher={Elsevier}
}

@article{bettarelli2023regional,
  title={A regional perspective to the study of affective polarization},
  author={Bettarelli, Luca and Reiljan, Andres and Van Haute, Emilie},
  journal={European Journal of Political Research},
  volume={62},
  number={2},
  pages={645--659},
  year={2023},
  publisher={Wiley Online Library}
}

@article{iyengar2012affect,
  title={Affect, not ideology: A social identity perspective on polarization},
  author={Iyengar, Shanto and Sood, Gaurav and Lelkes, Yphtach},
  journal={Public opinion quarterly},
  volume={76},
  number={3},
  pages={405--431},
  year={2012},
  publisher={Oxford University Press US}
}

@article{fiorina2008polarization,
  title={Polarization in the American public: Misconceptions and misreadings},
  author={Fiorina, Morris P and Abrams, Samuel A and Pope, Jeremy C},
  journal={The Journal of Politics},
  volume={70},
  number={2},
  pages={556--560},
  year={2008},
  publisher={Cambridge University Press New York, USA}
}

@article{abramowitz2010disappearing,
  title={The Disappearing Center: Engaged Citizens},
  author={Abramowitz, Alan I},
  journal={Polarization, and},
  year={2010}
}

@book{levendusky2009partisan,
  title={The partisan sort: How liberals became Democrats and conservatives became Republicans},
  author={Levendusky, Matthew},
  year={2009},
  publisher={University of Chicago Press}
}

@article{abramowitz2008polarization,
  title={Is polarization a myth?},
  author={Abramowitz, Alan I and Saunders, Kyle L},
  journal={The Journal of Politics},
  volume={70},
  number={2},
  pages={542--555},
  year={2008},
  publisher={Cambridge University Press New York, USA}
}

@book{fiorina2011culture,
  title={Culture war?},
  author={Fiorina, Morris P and Abrams, Samuel J and Pope, Jeremy C},
  year={2011},
  publisher={Longman}
}

@article{lelkes2016mass,
  title={Mass polarization: Manifestations and measurements},
  author={Lelkes, Yphtach},
  journal={Public Opinion Quarterly},
  volume={80},
  number={S1},
  pages={392--410},
  year={2016},
  publisher={Oxford University Press US}
}

@article{hasell2016partisan,
  title={Partisan provocation: The role of partisan news use and emotional responses in political information sharing in social media},
  author={Hasell, Ariel and Weeks, Brian E},
  journal={Human Communication Research},
  volume={42},
  number={4},
  pages={641--661},
  year={2016},
  publisher={Oxford University Press Oxford, UK}
}

@article{wischnewski2021shareworthiness,
  title={Shareworthiness and motivated reasoning in hyper-partisan news sharing behavior on Twitter},
  author={Wischnewski, Magdalena and Bruns, Axel and Keller, Tobias},
  journal={Digital Journalism},
  volume={9},
  number={5},
  pages={549--570},
  year={2021},
  publisher={Taylor \& Francis}
}

@article{kalogeropoulos2017shares,
  title={Who shares and comments on news?: A cross-national comparative analysis of online and social media participation},
  author={Kalogeropoulos, Antonis and Negredo, Samuel and Picone, Ike and Nielsen, Rasmus Kleis},
  journal={Social media+ society},
  volume={3},
  number={4},
  pages={2056305117735754},
  year={2017},
  publisher={SAGE Publications Sage UK: London, England}
}

@article{carpentras2023polarization,
  title={How polarization extends to new topics: An agent-based model derived from experimental data},
  author={Carpentras, Dino and Lueders, Adrian and Maher, Paul J and O'Reilly, Caoimhe and Quayle, Michael},
  journal={Journal of Artificial Societies and Social Simulation},
  volume={26},
  number={3},
  year={2023}
}

@article{dellaposta2015liberals,
  title={Why do liberals drink lattes?},
  author={DellaPosta, Daniel and Shi, Yongren and Macy, Michael},
  journal={American Journal of Sociology},
  volume={120},
  number={5},
  pages={1473--1511},
  year={2015},
  publisher={University of Chicago Press Chicago, IL}
}

@article{banisch2019opinion,
  title={Opinion polarization by learning from social feedback},
  author={Banisch, Sven and Olbrich, Eckehard},
  journal={The Journal of Mathematical Sociology},
  volume={43},
  number={2},
  pages={76--103},
  year={2019},
  publisher={Taylor \& Francis}
}

@article{tornberg2021modeling,
  title={Modeling the emergence of affective polarization in the social media society},
  author={T{\"o}rnberg, Petter and Andersson, Claes and Lindgren, Kristian and Banisch, Sven},
  journal={PLoS One},
  volume={16},
  number={10},
  pages={e0258259},
  year={2021},
  publisher={Public Library of Science San Francisco, CA USA}
}

@article{martin2023multipolar,
  title={Multipolar social systems: Measuring polarization beyond dichotomous contexts},
  author={Martin-Gutierrez, Samuel and Losada, Juan C and Benito, Rosa M},
  journal={Chaos, Solitons \& Fractals},
  volume={169},
  pages={113244},
  year={2023},
  publisher={Elsevier}
}

@article{fiorina1980decline,
  title={The decline of collective responsibility in American politics},
  author={Fiorina, Morris P},
  journal={Daedalus},
  pages={25--45},
  year={1980},
  publisher={JSTOR}
}

@article{page1993rational,
  title={The rational public and democracy},
  author={Page, Benjamin and Shapiro, Robert Y},
  journal={Reconsidering the democratic public},
  pages={35--64},
  year={1993},
  publisher={Pennsylvania State University Press University Park, PA}
}

@book{sunstein2006infotopia,
  title={Infotopia: How many minds produce knowledge},
  author={Sunstein, Cass R},
  year={2006},
  publisher={Oxford University Press}
}

@book{diamond1999developing,
  title={Developing democracy: Toward consolidation},
  author={Diamond, Larry},
  year={1999},
  publisher={JHU press}
}

@book{dahl2008polyarchy,
  title={Polyarchy: Participation and opposition},
  author={Dahl, Robert A},
  year={2008},
  publisher={Yale university press}
}

@Article{igraph,
    title = {The igraph software package for complex network research},
    author = {Gabor Csardi and Tamas Nepusz},
    journal = {InterJournal},
    volume = {Complex Systems},
    pages = {1695},
    year = {2006},
    url = {https://igraph.org},
  }

@article{cho2009percolation,
  title={Percolation transitions in scale-free networks under the Achlioptas process},
  author={Cho, Young Sul and Kim, Jin Seop and Park, Juyong and Kahng, Byungnam and Kim, Doochul},
  journal={Physical review letters},
  volume={103},
  number={13},
  pages={135702},
  year={2009},
  publisher={APS}
}

@article{chung2002connected,
  title={Connected components in random graphs with given expected degree sequences},
  author={Chung, Fan and Lu, Linyuan},
  journal={Annals of combinatorics},
  volume={6},
  number={2},
  pages={125--145},
  year={2002},
  publisher={Springer}
}

@article{goh2001universal,
  title={Universal behavior of load distribution in scale-free networks},
  author={Goh, K-I and Kahng, Byungnam and Kim, Doochul},
  journal={Physical review letters},
  volume={87},
  number={27},
  pages={278701},
  year={2001},
  publisher={APS}
}

@article{vasconcelos2021segregation,
  title={Segregation and clustering of preferences erode socially beneficial coordination},
  author={Vasconcelos, V{\'\i}tor V and Constantino, Sara M and Dannenberg, Astrid and Lumkowsky, Marcel and Weber, Elke and Levin, Simon},
  journal={Proceedings of the National Academy of Sciences},
  volume={118},
  number={50},
  pages={e2102153118},
  year={2021},
  publisher={National Acad Sciences}
}

@article{tokita2021polarized,
  title={Polarized information ecosystems can reorganize social networks via information cascades},
  author={Tokita, Christopher K and Guess, Andrew M and Tarnita, Corina E},
  journal={Proceedings of the National Academy of Sciences},
  volume={118},
  number={50},
  pages={e2102147118},
  year={2021},
  publisher={National Acad Sciences}
}

@article{macy2021polarization,
  title={Polarization and tipping points},
  author={Macy, Michael W and Ma, Manqing and Tabin, Daniel R and Gao, Jianxi and Szymanski, Boleslaw K},
  journal={Proceedings of the National Academy of Sciences},
  volume={118},
  number={50},
  pages={e2102144118},
  year={2021},
  publisher={National Acad Sciences}
}

@article{leonard2021nonlinear,
  title={The nonlinear feedback dynamics of asymmetric political polarization},
  author={Leonard, Naomi Ehrich and Lipsitz, Keena and Bizyaeva, Anastasia and Franci, Alessio and Lelkes, Yphtach},
  journal={Proceedings of the National Academy of Sciences},
  volume={118},
  number={50},
  pages={e2102149118},
  year={2021},
  publisher={National Acad Sciences}
}

@article{mas2013differentiation,
  title={Differentiation without distancing. Explaining bi-polarization of opinions without negative influence},
  author={M{\"a}s, Michael and Flache, Andreas},
  journal={PloS one},
  volume={8},
  number={11},
  pages={e74516},
  year={2013},
  publisher={Public Library of Science San Francisco, USA}
}

@article{macy2003polarization,
  title={Polarization in dynamic networks: A Hopfield model of emergent structure},
  author={Macy, Michael W and Kitts, James A and Flache, Andreas and Benard, Steve},
  year={2003},
  publisher={na}
}

@article{baldassarri2007dynamics,
  title={Dynamics of political polarization},
  author={Baldassarri, Delia and Bearman, Peter},
  journal={American sociological review},
  volume={72},
  number={5},
  pages={784--811},
  year={2007},
  publisher={Sage Publications Sage CA: Los Angeles, CA}
}

@incollection{axelrod1997advancing,
  title={Advancing the art of simulation in the social sciences},
  author={Axelrod, Robert},
  booktitle={Simulating social phenomena},
  pages={21--40},
  year={1997},
  publisher={Springer}
}

@article{knobloch2020confirmation,
  title={Confirmation bias, ingroup bias, and negativity bias in selective exposure to political information},
  author={Knobloch-Westerwick, Silvia and Mothes, Cornelia and Polavin, Nick},
  journal={Communication research},
  volume={47},
  number={1},
  pages={104--124},
  year={2020},
  publisher={Sage Publications Sage CA: Los Angeles, CA}
}

@article{bail2018exposure,
  title={Exposure to opposing views on social media can increase political polarization},
  author={Bail, Christopher A and Argyle, Lisa P and Brown, Taylor W and Bumpus, John P and Chen, Haohan and Hunzaker, MB Fallin and Lee, Jaemin and Mann, Marcus and Merhout, Friedolin and Volfovsky, Alexander},
  journal={Proceedings of the National Academy of Sciences},
  volume={115},
  number={37},
  pages={9216--9221},
  year={2018},
  publisher={National Acad Sciences}
}

@article{macy2002factors,
  title={From factors to actors: Computational sociology and agent-based modeling},
  author={Macy, Michael W and Willer, Robert},
  journal={Annual review of sociology},
  volume={28},
  number={1},
  pages={143--166},
  year={2002},
  publisher={Annual Reviews 4139 El Camino Way, PO Box 10139, Palo Alto, CA 94303-0139, USA}
}

@article{bonabeau2002agent,
  title={Agent-based modeling: Methods and techniques for simulating human systems},
  author={Bonabeau, Eric},
  journal={Proceedings of the national academy of sciences},
  volume={99},
  number={suppl\_3},
  pages={7280--7287},
  year={2002},
  publisher={National Acad Sciences}
}

@book{epstein1996growing,
  title={Growing artificial societies: social science from the bottom up},
  author={Epstein, Joshua M and Axtell, Robert},
  year={1996},
  publisher={Brookings Institution Press}
}

@article{schelling1971dynamic,
  title={Dynamic models of segregation},
  author={Schelling, Thomas C},
  journal={Journal of mathematical sociology},
  volume={1},
  number={2},
  pages={143--186},
  year={1971},
  publisher={Taylor \& Francis}
}

@article{neumann1966theory,
  title={Theory of self-reproducing automata},
  author={Neumann, J von},
  journal={Edited by Arthur W. Burks},
  year={1966}
}

@article{tornberg2022digital,
  title={How digital media drive affective polarization through partisan sorting},
  author={T{\"o}rnberg, Petter},
  journal={Proceedings of the National Academy of Sciences},
  volume={119},
  number={42},
  pages={e2207159119},
  year={2022},
  publisher={National Acad Sciences}
}

@article{garrett2009echo,
  title={Echo chambers online?: Politically motivated selective exposure among Internet news users},
  author={Garrett, R Kelly},
  journal={Journal of computer-mediated communication},
  volume={14},
  number={2},
  pages={265--285},
  year={2009},
  publisher={Oxford University Press Oxford, UK}
}

@article{wojcieszak2021echo,
  title={Echo chambers revisited: The (overwhelming) sharing of in-group politicians, pundits and media on Twitter},
  author={Wojcieszak, Magdalena and Casas, Andreu and Yu, Xudong and Nagler, Jonathan and Tucker, Joshua A},
  journal={OSF},
  volume={10},
  year={2021}
}

@article{madsen2018large,
  title={Large networks of rational agents form persistent echo chambers},
  author={Madsen, Jens Koed and Bailey, Richard M and Pilditch, Toby D},
  journal={Scientific reports},
  volume={8},
  number={1},
  pages={12391},
  year={2018},
  publisher={Nature Publishing Group UK London}
}

@article{barbera2015tweeting,
  title={Tweeting from left to right: Is online political communication more than an echo chamber?},
  author={Barber{\'a}, Pablo and Jost, John T and Nagler, Jonathan and Tucker, Joshua A and Bonneau, Richard},
  journal={Psychological science},
  volume={26},
  number={10},
  pages={1531--1542},
  year={2015},
  publisher={Sage Publications Sage CA: Los Angeles, CA}
}

@article{guess2018avoiding,
  title={Avoiding the echo chamber about echo chambers},
  author={Guess, Andrew and Nyhan, Brendan and Lyons, Benjamin and Reifler, Jason},
  journal={Knight Foundation},
  volume={2},
  number={1},
  pages={1--25},
  year={2018}
}

@article{dubois2018echo,
  title={The echo chamber is overstated: the moderating effect of political interest and diverse media},
  author={Dubois, Elizabeth and Blank, Grant},
  journal={Information, communication \& society},
  volume={21},
  number={5},
  pages={729--745},
  year={2018},
  publisher={Taylor \& Francis}
}

@book{sunstein2018republic,
  title={\# Republic: Divided democracy in the age of social media},
  author={Sunstein, Cass},
  year={2018},
  publisher={Princeton university press}
}

@article{lelkes2017hostile,
  title={The hostile audience: The effect of access to broadband internet on partisan affect},
  author={Lelkes, Yphtach and Sood, Gaurav and Iyengar, Shanto},
  journal={American Journal of Political Science},
  volume={61},
  number={1},
  pages={5--20},
  year={2017},
  publisher={Wiley Online Library}
}

@article{allcott2020welfare,
  title={The welfare effects of social media},
  author={Allcott, Hunt and Braghieri, Luca and Eichmeyer, Sarah and Gentzkow, Matthew},
  journal={American Economic Review},
  volume={110},
  number={3},
  pages={629--676},
  year={2020},
  publisher={American Economic Association 2014 Broadway, Suite 305, Nashville, TN 37203}
}

@article{chen2018effect,
  title={The effect of partisanship and political advertising on close family ties},
  author={Chen, M Keith and Rohla, Ryne},
  journal={Science},
  volume={360},
  number={6392},
  pages={1020--1024},
  year={2018},
  publisher={American Association for the Advancement of Science}
}

@article{huber2017political,
  title={Political homophily in social relationships: Evidence from online dating behavior},
  author={Huber, Gregory A and Malhotra, Neil},
  journal={The Journal of Politics},
  volume={79},
  number={1},
  pages={269--283},
  year={2017},
  publisher={University of Chicago Press Chicago, IL}
}

@article{kingzette2021affective,
  title={How affective polarization undermines support for democratic norms},
  author={Kingzette, Jon and Druckman, James N and Klar, Samara and Krupnikov, Yanna and Levendusky, Matthew and Ryan, John Barry},
  journal={Public Opinion Quarterly},
  volume={85},
  number={2},
  pages={663--677},
  year={2021},
  publisher={Oxford University Press}
}

@article{garzia2023affective,
  title={Affective Polarization in Comparative and Longitudinal Perspective},
  author={Garzia, Diego and Ferreira da Silva, Frederico and Maye, Simon},
  journal={Public Opinion Quarterly},
  volume={87},
  number={1},
  pages={219--231},
  year={2023},
  publisher={Oxford University Press}
}

@article{garzia2021negative,
  title={Negative personalization and voting behavior in 14 parliamentary democracies, 1961--2018},
  author={Garzia, Diego and da Silva, Frederico Ferreira},
  journal={Electoral Studies},
  volume={71},
  pages={102300},
  year={2021},
  publisher={Elsevier}
}

@article{boxell2022cross,
  title={Cross-country trends in affective polarization},
  author={Boxell, Levi and Gentzkow, Matthew and Shapiro, Jesse M},
  journal={Review of Economics and Statistics},
  pages={1--60},
  year={2022},
  publisher={MIT Press One Rogers Street, Cambridge, MA 02142-1209, USA journals-info~…}
}

@article{reiljan2020fear,
  title={‘Fear and loathing across party lines’(also) in Europe: Affective polarisation in European party systems},
  author={Reiljan, Andres},
  journal={European journal of political research},
  volume={59},
  number={2},
  pages={376--396},
  year={2020},
  publisher={Wiley Online Library}
}

@book{gidron2020american,
  title={American affective polarization in comparative perspective},
  author={Gidron, Noam and Adams, James and Horne, Will},
  year={2020},
  publisher={Cambridge University Press}
}

@article{kingzette2021you,
  title={Who do you loathe? Feelings toward politicians vs. ordinary people in the opposing party},
  author={Kingzette, Jon},
  journal={Journal of Experimental Political Science},
  volume={8},
  number={1},
  pages={75--84},
  year={2021},
  publisher={Cambridge University Press}
}

@article{druckman2021affective,
  title={Affective polarization, local contexts and public opinion in America},
  author={Druckman, James N and Klar, Samara and Krupnikov, Yanna and Levendusky, Matthew and Ryan, John Barry},
  journal={Nature human behaviour},
  volume={5},
  number={1},
  pages={28--38},
  year={2021},
  publisher={Nature Publishing Group UK London}
}

@article{druckman2019we,
  title={What do we measure when we measure affective polarization?},
  author={Druckman, James N and Levendusky, Matthew S},
  journal={Public Opinion Quarterly},
  volume={83},
  number={1},
  pages={114--122},
  year={2019},
  publisher={Oxford University Press UK}
}

@article{druckman2022mis,
  title={(Mis) estimating affective polarization},
  author={Druckman, James N and Klar, Samara and Krupnikov, Yanna and Levendusky, Matthew and Ryan, John Barry},
  journal={The Journal of Politics},
  volume={84},
  number={2},
  pages={1106--1117},
  year={2022},
  publisher={The University of Chicago Press Chicago, IL}
}

@article{gimpel2003partisan,
  title={Partisan Hearts and Minds: Political Parties and the Social Identities of Voters. By Donald Green, Bradley Palmquist, and Eric Schickler. New Haven, CT: Yale University Press, 2002. 272p. \$35.00},
  author={Gimpel, Jim},
  journal={Perspectives on Politics},
  volume={1},
  number={3},
  pages={606--607},
  year={2003},
  publisher={Cambridge University Press}
}

@article{iyengar2019origins,
  title={The origins and consequences of affective polarization in the United States},
  author={Iyengar, Shanto and Lelkes, Yphtach and Levendusky, Matthew and Malhotra, Neil and Westwood, Sean J},
  journal={Annual review of political science},
  volume={22},
  pages={129--146},
  year={2019},
  publisher={Annual Reviews}
}

@article{johnston2006party,
  title={Party identification: Unmoved mover or sum of preferences?},
  author={Johnston, Richard},
  journal={Annu. Rev. Polit. Sci.},
  volume={9},
  pages={329--351},
  year={2006},
  publisher={Annual Reviews}
}

@book{campbell1980american,
  title={The american voter},
  author={Campbell, Angus},
  year={1980},
  publisher={University of Chicago Press}
}

@article{lazarsfeld1954friendship,
  title={Friendship as a social process: A substantive and methodological analysis},
  author={Lazarsfeld, Paul F and Merton, Robert K and others},
  journal={Freedom and control in modern society},
  volume={18},
  number={1},
  pages={18--66},
  year={1954},
  publisher={New York, Van Nostrand}
}

@article{finkel2020political,
  title={Political sectarianism in America},
  author={Finkel, Eli J and Bail, Christopher A and Cikara, Mina and Ditto, Peter H and Iyengar, Shanto and Klar, Samara and Mason, Lilliana and McGrath, Mary C and Nyhan, Brendan and Rand, David G and others},
  journal={Science},
  volume={370},
  number={6516},
  pages={533--536},
  year={2020},
  publisher={American Association for the Advancement of Science}
}

@article{lelkes2017limits,
  title={The limits of partisan prejudice},
  author={Lelkes, Yphtach and Westwood, Sean J},
  journal={The Journal of Politics},
  volume={79},
  number={2},
  pages={485--501},
  year={2017},
  publisher={University of Chicago Press Chicago, IL}
}

@article{frimer2017liberals,
  title={Liberals and conservatives are similarly motivated to avoid exposure to one another's opinions},
  author={Frimer, Jeremy A and Skitka, Linda J and Motyl, Matt},
  journal={Journal of Experimental Social Psychology},
  volume={72},
  pages={1--12},
  year={2017},
  publisher={Elsevier}
}

@book{mason2018uncivil,
  title={Uncivil agreement: How politics became our identity},
  author={Mason, Lilliana},
  year={2018},
  publisher={University of Chicago Press}
}

@misc{newman2024reuters,
  title={Reuters Institute digital news report 2024 (Report of the Reuters Institute for the Study of Journalism)},
  author={Newman, Nic and Fletcher, Richard and Robertson, Craig T and Arguedas, Amy Ross and Nielsen, Rasmus K},
  year={2024},
  publisher={Oxford, UK: University of Oxford}
}

@article{van2017political,
  title={Political communication in a high-choice media environment: a challenge for democracy?},
  author={Van Aelst, Peter and Str{\"o}mb{\"a}ck, Jesper and Aalberg, Toril and Esser, Frank and De Vreese, Claes and Matthes, J{\"o}rg and Hopmann, David and Salgado, Susana and Hub{\'e}, Nicolas and St{\k{e}}pi{\'n}ska, Agnieszka and others},
  journal={Annals of the International Communication Association},
  volume={41},
  number={1},
  pages={3--27},
  year={2017},
  publisher={Taylor \& Francis}
}

@article{tucker2018social,
  title={Social media, political polarization, and political disinformation: A review of the scientific literature},
  author={Tucker, Joshua A and Guess, Andrew and Barber{\'a}, Pablo and Vaccari, Cristian and Siegel, Alexandra and Sanovich, Sergey and Stukal, Denis and Nyhan, Brendan},
  journal={Political polarization, and political disinformation: a review of the scientific literature (March 19, 2018)},
  year={2018}
}

@misc{stroud2011niche,
  title={Niche news: The politics of news choice},
  author={Stroud, Natalie Jomini},
  year={2011},
  publisher={Oxford University Press}
}

@article{kobayashi2024partisan,
  title={Is partisan selective exposure an American peculiarity? A comparative study of news browsing behaviors in the United States, Japan, and Hong Kong},
  author={Kobayashi, Tetsuro and Zhang, Zhifan and Liu, Ling},
  journal={Communication Research},
  pages={00936502241289109},
  year={2024},
  publisher={SAGE Publications Sage CA: Los Angeles, CA}
}

@article{fletcher2018people,
  title={Are people incidentally exposed to news on social media? A comparative analysis},
  author={Fletcher, Richard and Nielsen, Rasmus Kleis},
  journal={New media \& society},
  volume={20},
  number={7},
  pages={2450--2468},
  year={2018},
  publisher={Sage Publications Sage UK: London, England}
}

@article{weeks2017incidental,
  title={Incidental exposure, selective exposure, and political information sharing: Integrating online exposure patterns and expression on social media},
  author={Weeks, Brian E and Lane, Daniel S and Kim, Dam Hee and Lee, Slgi S and Kwak, Nojin},
  journal={Journal of computer-mediated communication},
  volume={22},
  number={6},
  pages={363--379},
  year={2017},
  publisher={Oxford University Press Oxford, UK}
}

@article{kubin2021role,
  title={The role of (social) media in political polarization: a systematic review},
  author={Kubin, Emily and Von Sikorski, Christian},
  journal={Annals of the International Communication Association},
  volume={45},
  number={3},
  pages={188--206},
  year={2021},
  publisher={Taylor \& Francis}
}

@book{settle2018frenemies,
  title={Frenemies: How social media polarizes America},
  author={Settle, Jaime E},
  year={2018},
  publisher={Cambridge University Press}
}

@article{overgaard2024perceiving,
  title={Perceiving affective polarization in the United States: How social media shape meta-perceptions and affective polarization},
  author={Overgaard, Christian Staal Bruun},
  journal={Social Media+ Society},
  volume={10},
  number={1},
  pages={20563051241232662},
  year={2024},
  publisher={SAGE Publications Sage UK: London, England}
}

@article{quattrociocchi2016echo,
  title={Echo chambers on Facebook},
  author={Quattrociocchi, Walter and Scala, Antonio and Sunstein, Cass R},
  journal={Available at SSRN 2795110},
  year={2016}
}

@article{interian2023network,
  title={Network polarization, filter bubbles, and echo chambers: An annotated review of measures and reduction methods},
  author={Interian, Ruben and G. Marzo, Rusl{\'a}n and Mendoza, Isela and Ribeiro, Celso C},
  journal={International Transactions in Operational Research},
  volume={30},
  number={6},
  pages={3122--3158},
  year={2023},
  publisher={Wiley Online Library}
}

@article{figa2022through,
  title={Through the newsfeed glass: Rethinking filter bubbles and echo chambers},
  author={Fig{\`a} Talamanca, Giacomo and Arfini, Selene},
  journal={Philosophy \& Technology},
  volume={35},
  number={1},
  pages={20},
  year={2022},
  publisher={Springer}
}

@article{guess2020does,
  title={Does counter-attitudinal information cause backlash? Results from three large survey experiments},
  author={Guess, Andrew and Coppock, Alexander},
  journal={British Journal of Political Science},
  volume={50},
  number={4},
  pages={1497--1515},
  year={2020},
  publisher={Cambridge University Press}
}

@article{kim2019cross,
  title={How cross-cutting news exposure relates to candidate issue stance knowledge, political polarization, and participation: The moderating role of political sophistication},
  author={Kim, Yonghwan},
  journal={International Journal of Public Opinion Research},
  volume={31},
  number={4},
  pages={626--648},
  year={2019},
  publisher={Oxford University Press}
}

@article{lin2023effects,
  title={The effects of disagreement and unfriending on political polarization: a moderated-mediation model of cross-cutting discussion on affective polarization via unfriending contingent upon exposure to incivility},
  author={Lin, Han and Wang, Yi and Lee, Janggeun and Kim, Yonghwan},
  journal={Journal of Computer-Mediated Communication},
  volume={28},
  number={4},
  pages={zmad022},
  year={2023},
  publisher={Oxford University Press}
}

@article{zhu2024implications,
  title={Implications of online incidental and selective exposure for political emotions: Affective polarization during elections},
  author={Zhu, Qinfeng and Weeks, Brian E and Kwak, Nojin},
  journal={new media \& society},
  volume={26},
  number={1},
  pages={450--472},
  year={2024},
  publisher={SAGE Publications Sage UK: London, England}
}

@article{arendt2015toward,
  title={Toward a dose-response account of media priming},
  author={Arendt, Florian},
  journal={Communication Research},
  volume={42},
  number={8},
  pages={1089--1115},
  year={2015},
  publisher={Sage Publications Sage CA: Los Angeles, CA}
}

@article{lin2025exploring,
  title={Exploring the Paradox of Cross-Cutting Exposure and Affective Polarization: A Curvilinear Model Influenced by Political Ideology Strength},
  author={Lin, Han and Jiang, Xuejin and Lee, Janggeun and Wang, Yi and Kim, Yonghwan},
  journal={Media Psychology},
  pages={1--25},
  year={2025},
  publisher={Taylor \& Francis}
}

@article{wojcieszak2020can,
  title={Can interparty contact reduce affective polarization? A systematic test of different forms of intergroup contact},
  author={Wojcieszak, Magdalena and Warner, Benjamin R},
  journal={Political Communication},
  volume={37},
  number={6},
  pages={789--811},
  year={2020},
  publisher={Taylor \& Francis}
}

@article{chen2022effect,
  title={The effect of cross-cutting exposure on attitude change: examining the mediating role of response behaviors and the moderating role of openness to diversity and social network homogeneity},
  author={Chen, Hsuan-Ting and Ai, Minwei and Guo, Jing},
  journal={Asian Journal of Communication},
  volume={32},
  number={2},
  pages={93--110},
  year={2022},
  publisher={Taylor \& Francis}
}

@article{turner1979social,
  title={Social comparison and group interest in ingroup favouritism},
  author={Turner, John C and Brown, Rupert J and Tajfel, Henri},
  journal={European journal of social psychology},
  volume={9},
  number={2},
  pages={187--204},
  year={1979},
  publisher={Wiley Online Library}
}

@article{skytte2021dimensions,
  title={Dimensions of elite partisan polarization: Disentangling the effects of incivility and issue polarization},
  author={Skytte, Rasmus},
  journal={British Journal of Political Science},
  volume={51},
  number={4},
  pages={1457--1475},
  year={2021},
  publisher={Cambridge University Press}
}

@article{huddy2021reducing,
  title={Reducing affective polarization: Warm group relations or policy compromise?},
  author={Huddy, Leonie and Yair, Omer},
  journal={Political Psychology},
  volume={42},
  number={2},
  pages={291--309},
  year={2021},
  publisher={Wiley Online Library}
}

@article{horne2023way,
  title={The way we were: how histories of co-governance alleviate partisan hostility},
  author={Horne, Will and Adams, James and Gidron, Noam},
  journal={Comparative Political Studies},
  volume={56},
  number={3},
  pages={299--325},
  year={2023},
  publisher={SAGE Publications Sage CA: Los Angeles, CA}
}

@article{back2023elite,
  title={Elite communication and affective polarization among voters},
  author={B{\"a}ck, Hanna and Carroll, Royce and Renstr{\"o}m, Emma and Ryan, Alexander},
  journal={Electoral Studies},
  volume={84},
  pages={102639},
  year={2023},
  publisher={Elsevier}
}

@article{wagner2024elite,
  title={Elite Cooperation and Affective Polarization: Evidence From German Coalitions},
  author={Wagner, Markus and Harteveld, Eelco},
  journal={Political Studies},
  pages={00323217241300993},
  year={2024},
  publisher={SAGE Publications Sage UK: London, England}
}

@article{delli2001let,
  title={Let us infotain you: Politics in the new media age},
  author={Delli Carpini, Michael X and Williams, Bruce A},
  journal={Mediated politics: Communication in the future of democracy},
  number={14},
  pages={160--181},
  year={2001}
}

@article{APSA1950,
  author    = {American Political Science Association},
  title     = {Toward a More Responsible Two-Party System},
  journal   = {American Political Science Review},
  year      = {1950},
  volume    = {44},
  publisher = {American Political Science Association}
}

@article{reiljan2024patterns,
  title={Patterns of affective polarization toward parties and leaders across the democratic world},
  author={Reiljan, Andres and Garzia, Diego and Da Silva, Frederico Ferreira and Trechsel, Alexander H},
  journal={American Political Science Review},
  volume={118},
  number={2},
  pages={654--670},
  year={2024},
  publisher={Cambridge University Press}
}

@article{kreiss2024review,
  title={A review and provocation: On polarization and platforms},
  author={Kreiss, Daniel and McGregor, Shannon C},
  journal={New Media \& Society},
  volume={26},
  number={1},
  pages={556--579},
  year={2024},
  publisher={SAGE Publications Sage UK: London, England}
}

@article{barbera2015birds,
  title={Birds of the same feather tweet together: Bayesian ideal point estimation using Twitter data},
  author={Barber{\'a}, Pablo},
  journal={Political analysis},
  volume={23},
  number={1},
  pages={76--91},
  year={2015},
  publisher={Cambridge University Press}
}

@article{garimella2018quantifying,
  title={Quantifying controversy on social media},
  author={Garimella, Kiran and Morales, Gianmarco De Francisci and Gionis, Aristides and Mathioudakis, Michael},
  journal={ACM Transactions on Social Computing},
  volume={1},
  number={1},
  pages={1--27},
  year={2018},
  publisher={ACM New York, NY, USA}
}

@article{filsinger2024asymmetric,
  title={Asymmetric affective polarization regarding COVID-19 vaccination in six European countries},
  author={Filsinger, Maximilian and Freitag, Markus},
  journal={Scientific Reports},
  volume={14},
  number={1},
  pages={15919},
  year={2024},
  publisher={Nature Publishing Group UK London}
}
\appendix

\section{Perception Towards News Media: Survey Introduction}

The primary objective of this survey is to validate the key assumptions that underlie the proposed model. Specifically, the survey aims to test the following assumptions,

\begin{enumerate}
\item Individuals are more likely to disseminate news that aligns with their ideological perspectives.
\item Exposure to news content with opposing ideological perspectives may provoke feelings of animosity toward the opposing party.
\end{enumerate}






To achieve this objective, we conducted a crowd-sourced annotation task via Prolific between December 1 and December 10, 2024, recruiting 200 U.S.-based participants eligible to vote ($N_{Dem} = 97,~ N_{Rep} = 99$ and $N_{Ind} = 4$). Participants responded to five sets of questions, each randomly selected from a pool of 27 predefined sets focused on prominent U.S. media outlets, derived from a recent Pew Research Center report \parencite{jurkowitz2020us}. Each set of questions was designed to assess individuals' feelings of trust, affect, and likelihood of resharing. 

Trust in each outlet was measured using a 5-point Likert scale, ranging from 1 ("Not at all") to 5 ("A lot"), with an additional option for "Don't know/Can't say" to capture uncertainty. Emotional reactions to each outlet were evaluated using a "feeling thermometer," where scores ranged from 0 (most negative) to 100 (most positive). Participants were also asked to assess their likelihood of resharing news from each outlet on social media, utilizing another 5-point Likert scale, from 1 ("Very unlikely") to 5 ("Very likely"), with an option for "Don't know/Can't say." It is important to note that these responses were collected without exposing the participants to specific news content.

\begin{figure}[htp]
\centering
\includegraphics[width=\textwidth]{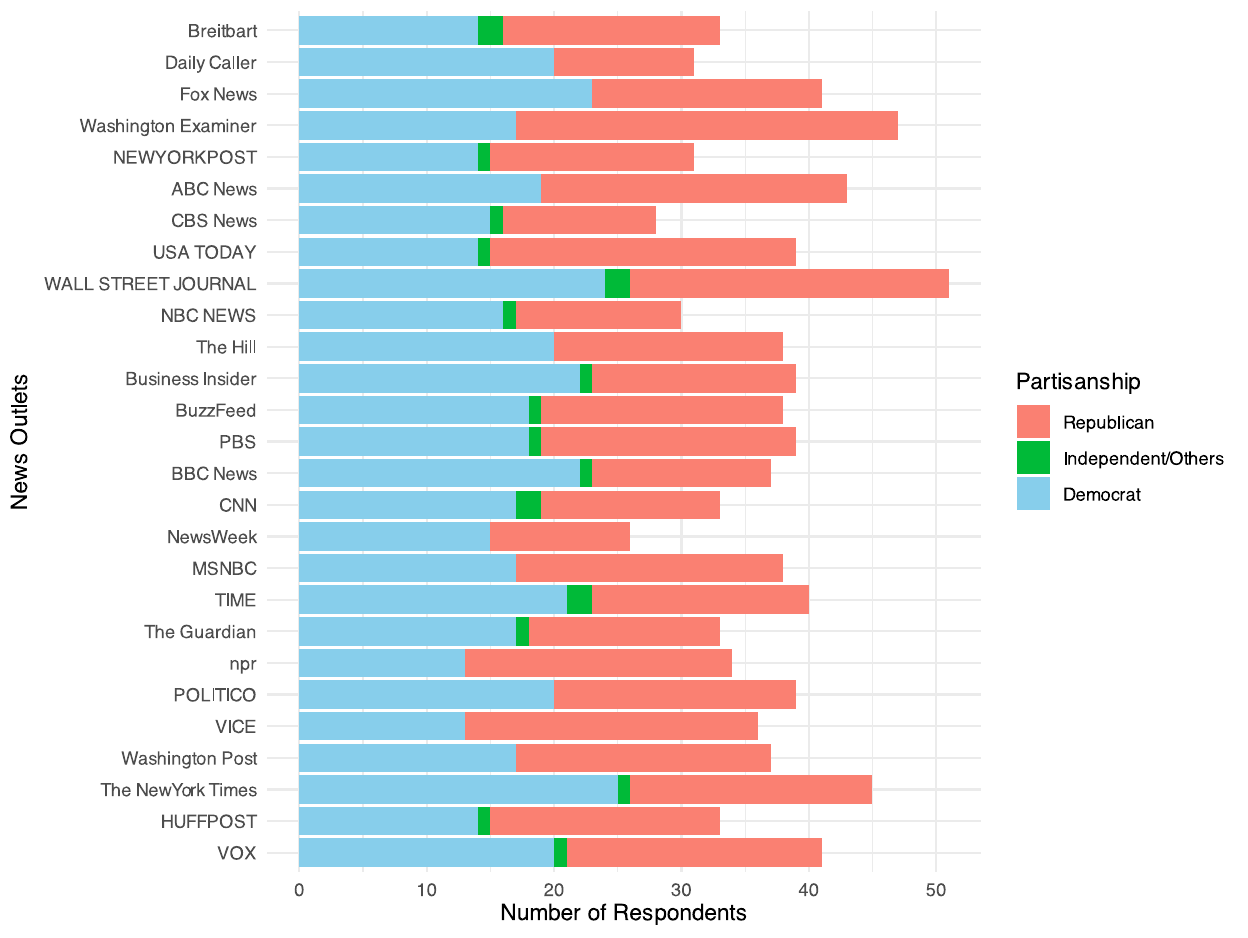}
\caption{Total number of respondents by party identification and media outlet\\
Note: The bars illustrate the total number of respondents -- Democrats, Republicans, and Independents -- who were presented with each media outlet. This includes both those who recognized and those who did not recognize the outlets presented to them.}
\label{fig6a}
\end{figure}

\begin{figure}[htp]
\centering
\includegraphics[width=\textwidth]{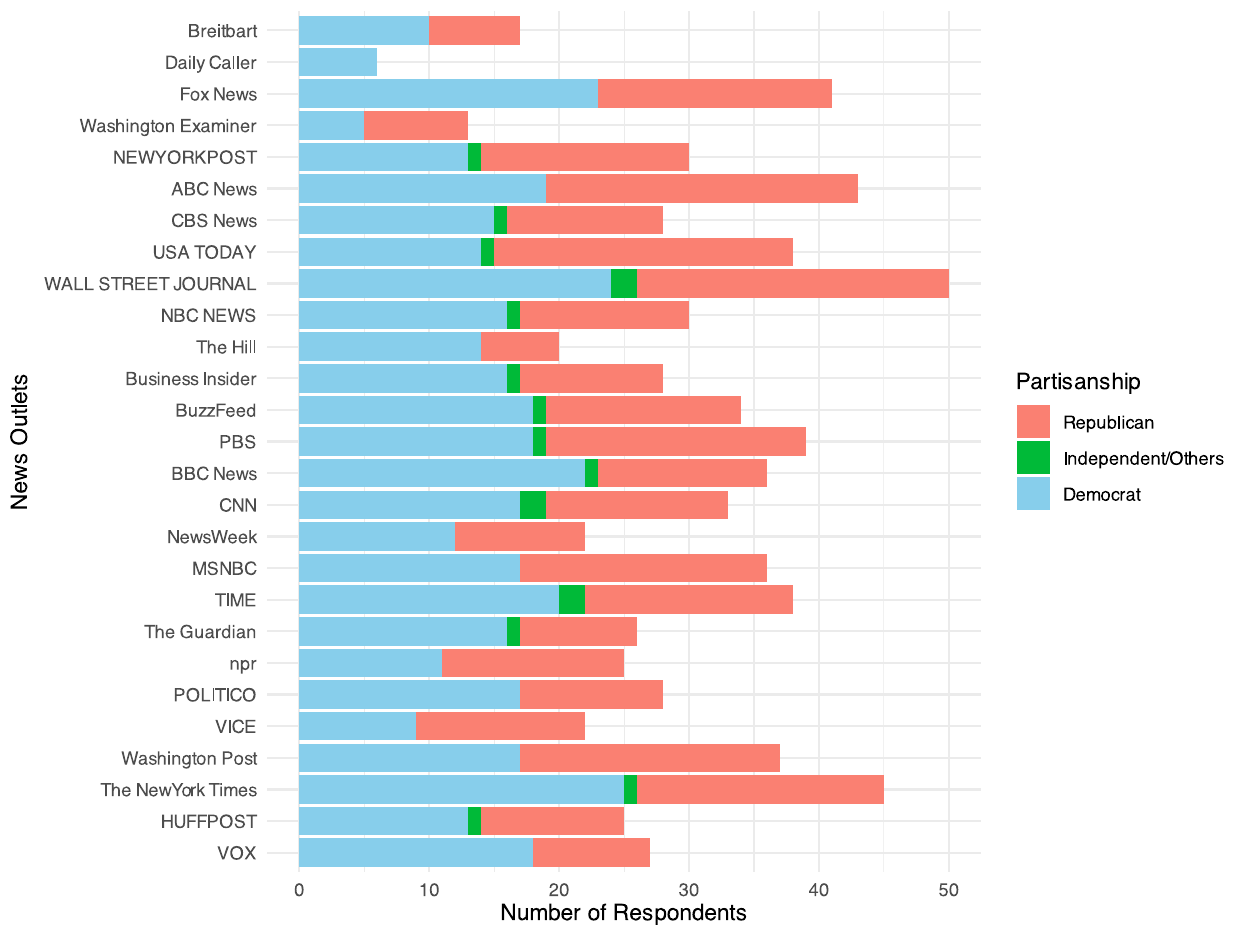}
\caption{Total number of respondents by party identification and media outlet recognition\\
Note: The bars illustrate the total number of respondents -- Democrats, Republicans, and Independents -- who were presented with each media outlet and recognized the same.}
\label{fig7a}
\end{figure}

Furthermore, participants were required to indicate whether they recognized each outlet and how frequently they consumed news from it, allowing for an assessment of their familiarity with the outlets in question. Demographic information, including gender, ethnicity, and political affiliation (Democrat or Republican), was also collected. The partisan profile of respondents to each of the 27 sets of questions can be seen in Figure \ref{fig6a}. The Figure \ref{fig7a} represents participants who responded that they recognised the outlet presented to them. Additionally, to ensure participant engagement, two attention checks were incorporated into the survey. Responses from participants who failed these checks were excluded from the final analysis. 


\section{Results}

\begin{figure}[htp]
\centering
\includegraphics[width=\textwidth]{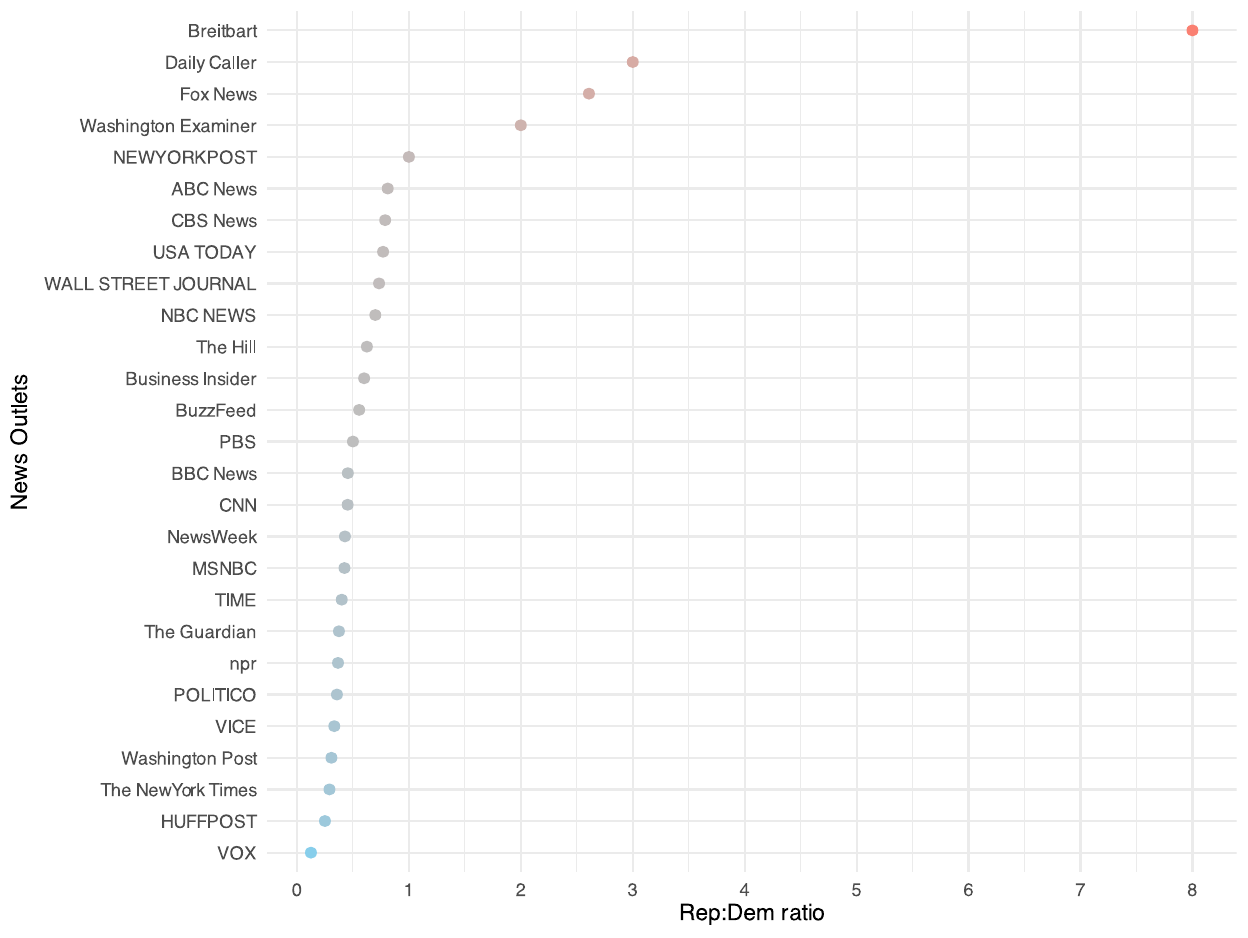}
\caption{Partisan bias of shortlisted media outlets based on viewers' partisan inclination\\
Note: The point values denote the Republican-to-Democrat (Rep:Dem) ratio, where lower values indicate viewership predominantly among Democrats, and higher values indicate a majority Republican viewership. Intermediate values indicate relatively balanced patronage from both partisan groups.}
\label{fig8a}
\end{figure}

We assess the partisan orientation of media outlets by analyzing the proportion of Republican and Democratic-leaning individuals who reported trusting each outlet for political and election news, following the methodology established in prior studies \parencite{jurkowitz2020us}. Specifically, we calculate the partisan leaning of each outlet by deriving the Republican-to-Democrat (Rep:Dem) ratio using the following formula,
\begin{equation}
\text{Rep:Dem} := \frac{\%\ of\ Republicans\
who\ trust \ the \ outlet}{\%\ of\ Democrats\ who\ trust \ the \ outlet} .
\end{equation}
Accordingly, lower values of the ratio indicate a Democratic leaning, whereas higher values suggest a Republican orientation of the respective media outlet (see Figure \ref{fig8a}).

\subsection{Trust}

\begin{figure}[htp]
\centering
\includegraphics[width=\textwidth]{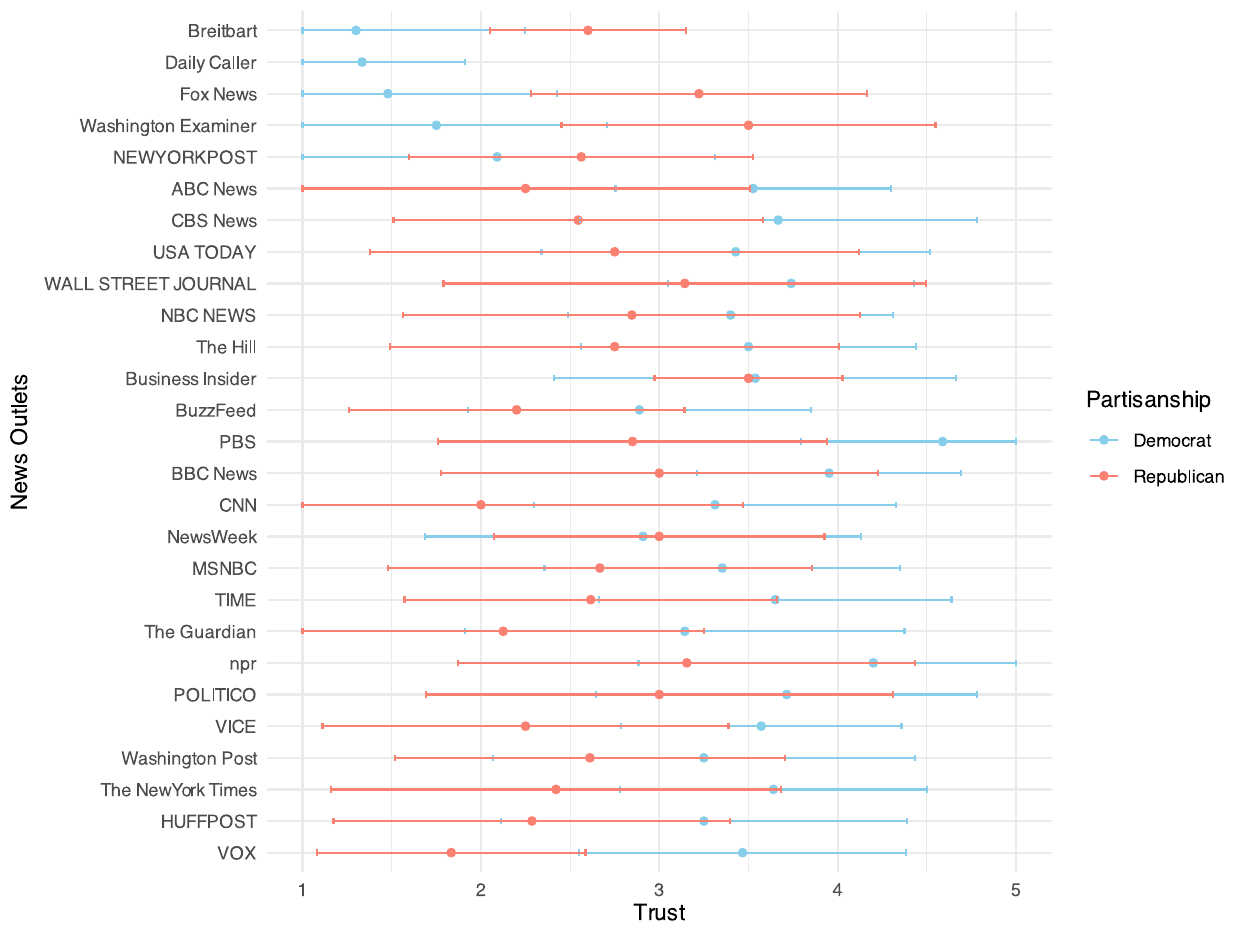}
\caption{Mean trust scores of media outlets by partisan affiliation\\
Note: The point values denote the mean values and the bars represent standard deviations of trust scores for each media outlet, aggregated from survey participants' responses. $Trust_{j(Dem)}$ is represented in blue, indicating Democrats' trust in the respective media outlet, while $Trust_{j(Rep)}$ is shown in red, representing Republicans' trust. Higher point values indicate greater aggregate trust towards the outlet among partisans. Media outlets are arranged in the increasing order of their $Rep:Dem$ ratio, with Republican-leaning outlets positioned at the top of the y-axis and Democrat-leaning outlets at the bottom. Consequently, outlets at the top are more trusted by Republicans, while those at the bottom are more trusted by Democrats.}
\label{fig9a}
\end{figure}

The mean trust scores for each media outlet, aggregated across all participants who recognized the respective outlet, are displayed in Figure \ref{fig9a}. The mean values corresponding to each outlet ($\text{Trust}_{j (Dem\ or\ Rep)}$) is computed using the formula below,
\begin{equation}
\text{Trust}_{j (Dem\ or\ Rep)}~:=~\frac{\sum\limits_{i=1}^{N_{Dem(or\ Rep)}} trust_{i}}{N_{Dem(or\ Rep)}},~ i= 1,2,\hdots, N_{Dem(or\ Rep)},~\forall j\leq|Outlets|.
\end{equation}
Where $N_{Dem(or\ Rep)}$ represents the total number of Democratic or Republican participants who recognized the media outlet presented to them, $trust_{i}$ denotes the respective trust score provided by each participant, and $|Outlets|$ represents the total number of media outlets being analyzed.

The trust scores assigned to media outlets by respondents reveal a clear partisan divide in perceptions of credibility. Specifically, the data demonstrate that Republicans and Democrats are more likely to trust media outlets they perceive as aligning with their political biases, while exhibiting lower trust in outlets perceived to favor opposing viewpoints.

\subsection{Information cascade}

\begin{figure}[htp]
\centering
\includegraphics[width=\textwidth]{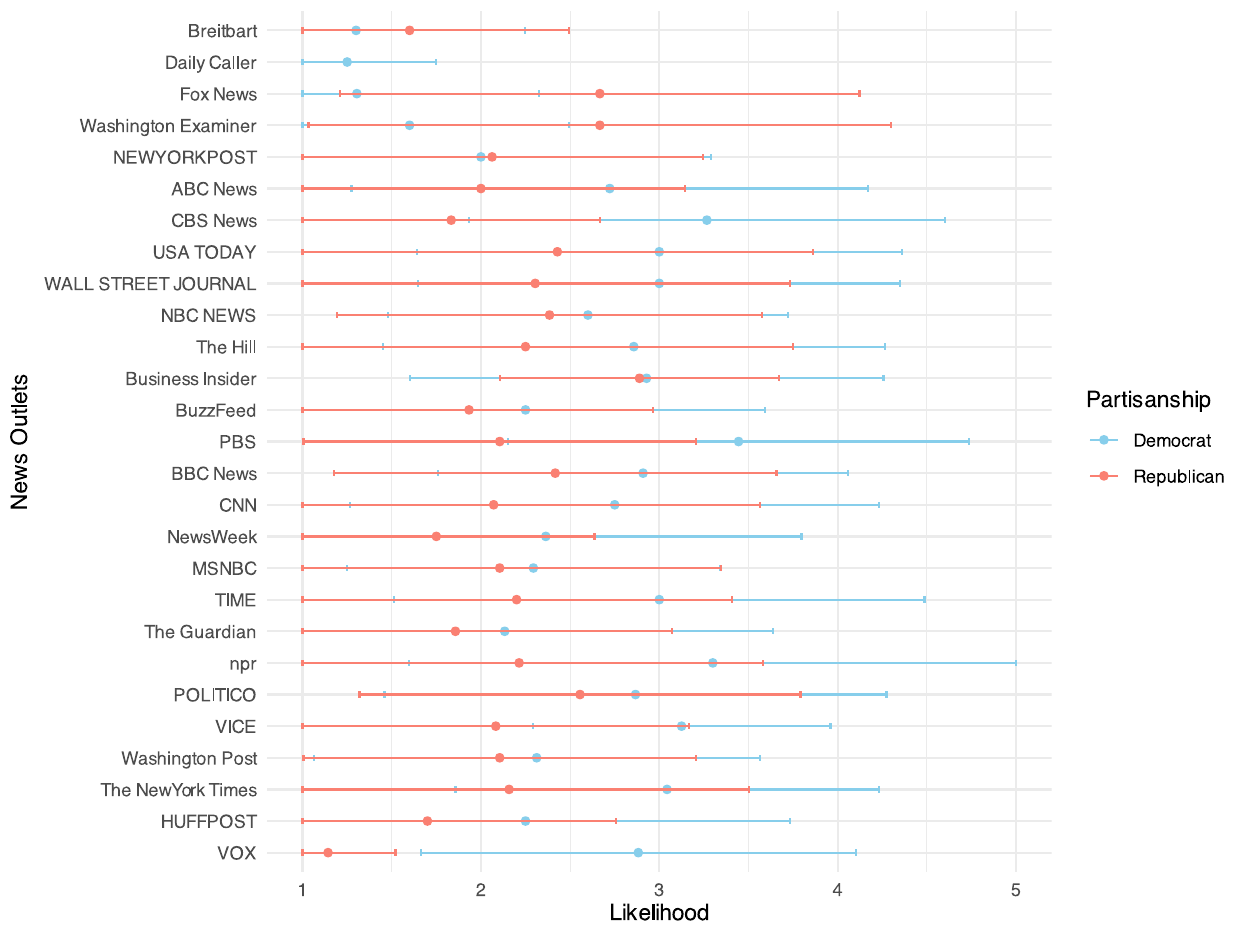}
\caption{Mean news sharing likelihood of media outlets by partisan affiliation\\
Note: The point values denote the mean values and the bars represent standard deviations of news sharing for each media outlet, aggregated from survey participants' responses. $Share_{j(Dem)}$ is represented in blue, indicating Democrats' likelihood of sharing news from the respective media outlet, while $Share_{j(Rep)}$ is shown in red, representing Republicans' likelihood of sharing news. Higher point values indicate greater likelihood of sharing news from the outlet among partisans. Media outlets are arranged in the increasing order of their $Rep:Dem$ ratio, with Republican-leaning outlets positioned at the top of the y-axis and Democrat-leaning outlets at the bottom. Consequently, news from outlets at the top of the plot is more likely to be shared by Republicans, while news from outlets at the bottom is more likely to be shared by Democrats.}
\label{fig10a}
\end{figure}

The mean values of the likelihood of information conveyed by the media outlet being shared by consumers across partisan groups are presented in Figure \ref{fig10a}. The mean likelihood values for each outlet ($\text{Share}_{j  (Dem\ or\ Rep)}$) is computed using the equation below,
\begin{equation}
\text{Share}_{j (Dem\ or\ Rep)}~:=~\frac{\sum\limits_{i=1}^{N_{Dem(or\ Rep)}} share_{i}}{N_{Dem(or\ Rep)}},~ i= 1,2,\hdots, N_{Dem(or\ Rep)},~\forall j\leq|Outlets|.
\end{equation}
Where $N_{Dem(or\ Rep)}$ represents the total number of Democratic or Republican participants who recognized the media outlet presented to them, $share_{i}$ denotes the likelihood of information being shared as indicated by each participant, and $|Outlets|$ represents the total number of media outlets analyzed.

Republicans generally reported a higher likelihood of sharing news from right-leaning outlets compared to Democrats. However, balanced or neutral outlets hold cross-partisan appeal in the context of news-sharing behavior; trends indicate that consumers across partisan groups are equally likely to share information published by these sources.





\subsection{Affect}

\begin{figure}[htp]
\centering
\includegraphics[width=\textwidth]{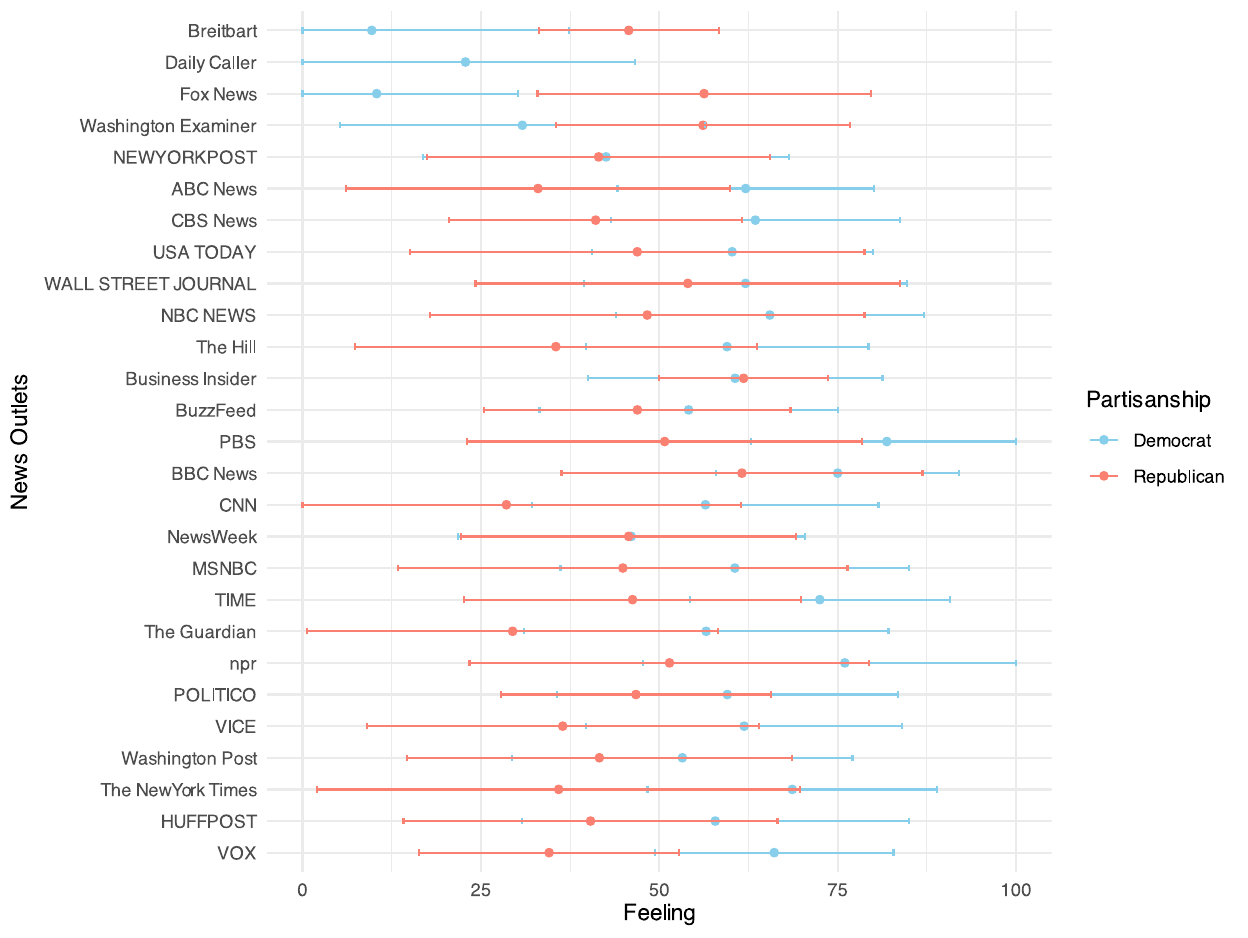}
\caption{Mean affective response to media outlets by partisan affiliation\\
Note: The point values denote the mean values and the bars represent standard deviations of affect for each media outlet, aggregated from survey participants' responses. $Affect_{j(Dem)}$ is represented in blue, indicating Democrats' affect towards the respective media outlet, while $Affect_{j(Rep)}$ is shown in red, representing Republicans' affect. Higher point values indicate feelings of warmth (or loyalty) towards the outlet among partisans. Media outlets are arranged in the increasing order of their $Rep:Dem$ ratio, with Republican-leaning outlets positioned at the top of the y-axis and Democrat-leaning outlets at the bottom. Accordingly, Republicans show greater loyalty toward outlets at the top of the plot, while Democrats express more loyal toward outlets at the bottom.}
\label{fig11a}
\end{figure}

The mean affect scores for each media outlet, aggregated across all participants who recognized the respective outlet, are displayed in Figure \ref{fig11a}. The mean values of affect corresponding to each outlet ($\text{Affect}_{j (Dem\ or\ Rep)}$) is computed using the formula below,
\begin{equation}
\text{Affect}_{j (Dem\ or\ Rep)}~:=~\frac{\sum\limits_{i=1}^{N_{Dem(or\ Rep)}} affect_{i}}{N_{Dem(or\ Rep)}},~ i= 1,2,\hdots, N_{Dem(or\ Rep)},~\forall j\leq|Outlets|.
\end{equation}
Where $N_{Dem(or\ Rep)}$ represents the total number of Democratic or Republican participants who recognized the media outlet presented to them, $affect_{i}$ denotes the respective affect provided by each participant, and $|Outlets|$ represents the total number of media outlets being analyzed.

The mean affect values for Republicans toward right-leaning outlets, such as the Washington Examiner, Fox News, and Breitbart, indicate significantly warmer or more positive sentiments compared to Democrats, who generally report colder or more negative feelings toward these outlets. Conversely, the mean affect values for left-leaning outlets, including VOX, VICE, TIME, The New York Times, and NPR, reveal that Democrats exhibit notably positive feelings, in contrast to Republicans, who display relatively colder sentiments toward these sources. For perceived neutral outlets, the affective responses appear to be mixed. Some outlets \textendash\ such as The Wall Street Journal, New York Post and Business Insider \textendash\ elicit similar levels of affect across partisan groups, whereas others \textendash\ such as ABC News, The Hill, NBC News and CBS News \textendash\ exhibit notable differences in affective sentiments between Republicans and Democrats.


This survey systematically evaluates individuals’ trust in media outlets, their affective responses, and their likelihood of resharing content. The findings confirm that individuals preferentially engage with news sources aligned with their ideological perspectives, while exposure to ideologically opposing content provokes varying affective reactions. Furthermore, the observed partisan patterns in media trust and information-sharing behaviors provide empirical support for the theoretical foundations of our proposed model, particularly the tendencies of individuals to not only trust media outlets that align with their partisan orientations, but also to disseminate news content from those sources.

\section{Questionnaire}
The survey questionnaire questionnaire can be found here: \url{https://osf.io/f9pzm/?view_only=441a2ed837da445fb7ac7b978655dcc3}.


\end{document}